\documentclass[prd,a4paper,nofootinbib,notitlepage,longbibliography,10pt
]{revtex4-1}
\usepackage{amsmath}
\usepackage{graphicx}
\usepackage{amsfonts}
\usepackage{amssymb}
\usepackage{amsthm}
	\renewcommand{\qedsymbol}{\rule{0.7em}{0.7em}}
\usepackage{slashed} 
\usepackage{color}
\usepackage{hyperref}
\usepackage{bbm,bm}
\usepackage{psfrag}
\usepackage{dsfont}
\usepackage{booktabs}
\usepackage[markup=nocolor]{changes}
\usepackage{slashed}
\usepackage{cleveref}
\usepackage[utf8]{inputenc}
\usepackage[american]{babel}
\usepackage{caption}
\usepackage{newfloat}
\usepackage{comment}

\DeclareFloatingEnvironment[fileext=lop]{Table}

\graphicspath{{./figures/}}

\definecolor{oucrimsonred}{rgb}{0.6, 0.0, 0.0}
\definecolor{persianblue}{rgb}{0.11, 0.22, 0.73}
\definecolor{forestgreen}{rgb}{0.13,0.35,0.13}
\hypersetup{colorlinks, citecolor=oucrimsonred, linkcolor=persianblue, urlcolor=oucrimsonred}
\newcommand{\be}{\begin{equation}}
\newcommand{\ee}{\end{equation}}
\newcommand{\bea}{\begin{eqnarray}}
\newcommand{\eea}{\end{eqnarray}}
\newcommand{\nn}{\nonumber}

\newcommand{\Lagr}{\mathcal{L}}

\newcommand{\rmc}{\mathrm{c}}
\newcommand{\rmt}{\mathrm{t}}
\newcommand{\rmb}{\mathrm{b}}
\newcommand{\rmL}{\mathrm{L}}
\newcommand{\rmR}{\mathrm{R}}
\newcommand{\rmY}{\mathrm{Y}}
\newcommand{\rmV}{\mathrm{V}}
\newcommand{\rmS}{\mathrm{S}}
\newcommand{\SM}{\mathrm{SM}}

\newcommand{\VLF}{\Psi}
\newcommand{\psiL}{\psi_L}
\newcommand{\psiV}{\psi_V}
\newcommand{\psiG}{\psi_G}
\newcommand{\psiU}{\psi_U}
\newcommand{\psiQ}{\psi_Q}
\newcommand{\ML}{M_L}
\newcommand{\MV}{M_V}
\newcommand{\MG}{M_G}
\newcommand{\MU}{M_U}
\newcommand{\MQ}{M_Q}
\newcommand{\psibarL}{\overline{\psi}_L}
\newcommand{\psibarV}{\overline{\psi}_V}
\newcommand{\psibarG}{\overline{\psi}_G}
\newcommand{\LL}{$L$}
\newcommand{\V}{$V$}
\newcommand{\G}{$G$}
\newcommand{\UU}{$U$}
\newcommand{\Q}{$Q$}
\newcommand{\LVG}{\rm LVG}
\newcommand{\UUQQ}{\rm 2U2Q}
\newcommand{\yV}{y_V}
\newcommand{\yt}{y_{\rm t}}
\newcommand{\ytM}{y_{\rm t10}}
\newcommand{\alphaV}{\alpha_{V}}
\newcommand{\alphat}{\alpha_{\rm t}}
\newcommand{\alphatm}{\alpha_{\rm t}}
\newcommand{\alphatM}{\alpha_{\rm t10}}
\newcommand{\alphaM}{\alpha_5}
\newcommand{\alphaQ}{\alpha_{Q}}
\newcommand{\yQ}{y_{Q}}
\newcommand{\calY}{\mathcal{Y}}
\newcommand{\MGUT}{$M_{\rm GUT}$}

\newcommand{\rmP}{\mathrm{P}}
\newcommand{\Tr}{\mathrm{Tr}}
\newcommand{\rmT}{\mathrm{T}}

\newcommand{\I}{\mathrm{i}}
\newcommand{\E}{\mathrm{e}}

\newcommand{\Nf}{N_{\mathrm{f}}}

\newcommand{\SU}{\mathrm{SU}}
\newcommand{\U}{\mathrm{U}}
\newcommand{\OO}{\mathrm{O}}
\newcommand{\rmF}{\mathrm{F}}

\newcommand{\de}{\partial}
\newcommand{\Eqref}[1]{Eq.~\eqref{#1}}
\newcommand{\Figref}[1]{Fig.~\ref{#1}}
\newcommand{\Secref}[1]{Sec.~\ref{#1}}
\newcommand{\Secrefs}[1]{Secs.~\ref{#1}}
\newcommand{\Appref}[1]{App.~\ref{#1}}
\newcommand{\Tabref}[1]{Tab.~\ref{#1}}
\newcommand{\Tabrefs}[1]{Tabs.~\ref{#1}}
\newcommand{\Citeref}[1]{\cite{#1}}
\newcommand{\Citerefs}[1]{\cite{#1}}
  {\left\lbrace\begin{array}{@{}l@{}}}%
  {\end{array}\right.}

\definecolor{ao(english)}{rgb}{0.0, 0.5, 0.0}
\definecolor{hg}{rgb}{0.8, 0, 0.5}

\begin{document}
\title[]{An asymptotically safe SU(5) GUT}
\date{December 4, 2020}
\author{M.\ Fabbrichesi$^{\dag}$}
\author{C.\ M.\ Nieto$^{\ast}$}
\author{A.\ Tonero$^{\ddag}$}
\author{A.\ Ugolotti$^{\bullet}$}
\affiliation{$^{\dag}$INFN, Sezione di Trieste, Via  Valerio 2, 34127 Trieste, Italy }
\affiliation{$^{\ast}$Universidad Industrial de Santander, Carrera 27 Calle 9, Bucaramanga, Santander, Colombia }
\affiliation{$^{\ddag}$ Ottawa-Carleton Institute for Physics, Carleton University, 1125 Colonel By Drive, Ottawa, Ontario K1S 5B6, Canada}
\affiliation{$^{\bullet}$Theoretisch-Physikalische Institut (TPI), Friedrich-Schiller-Universit{\"a}t,
Abbeanum, Fr{\"o}belstieg 1, 07743 Jena, Germany }

\begin{abstract}
\noindent We minimally extend the Standard Model field content by adding new vector-like fermions at the  TeV scale to allow  gauge coupling unification at a realistic scale.  We  embed the model into a $\SU(5)$ grand unified theory that is asymptotically safe and features an interacting fixed point for the gauge coupling. There are no Landau poles of the $\U(1)$ gauge  and Higgs couplings. Gauge, Yukawa and Higgs couplings are retraced from the fixed point and   matched at the grand unification scale to those of the  Standard Model rescaled up to the same energy. All couplings, their fixed point values and critical exponents always remain in the perturbative regime.
\end{abstract}

\maketitle 

\section{Motivations}
\label{sec:Motivations}

A quantum field theory is asymptotically safe if all its couplings reach a fixed point in the ultraviolet (UV) limit, as they run along the flow dictated by their renormalization group equations~\cite{Wilson:1973jj,Weinberg:1980gg}. The fixed point can be interacting or free (Gau\ss ian). In the latter case, asymptotic safety reduces to asymptotic freedom~\cite{Gross:1973id,Politzer:1973fx}. In both cases, we can say that the theory is UV complete because it is well behaved and predictive at all energies.

The Standard Model (SM) is not asymptotically safe because of the uncertain fate of the Higgs boson quartic coupling and the presence of the Landau pole in the $\U(1)_\rmY$ gauge coupling.
In particular, the latter divergence feeds back into the renormalization group (RG) flow of the quartic Higgs self-interaction inducing a Landau pole also in the scalar sector.
Furthermore, the Higgs quartic coupling---given the current experimental value of the top mass---becomes negative before the Planck scale, making the electroweak vacuum metastable~\cite{Isidori:2001bm}.
Quite in general, for a given cutoff scale and fixed value for the top mass, the Higgs mass has to exceed a lower bound in order to avoid the metastability issue of the scalar potential~\cite{Ellis:2009tp,EliasMiro:2011aa,Degrassi:2012ry,Alekhin:2012py,Masina:2012tz,Buttazzo:2013uya}. Although different mechanisms can be devised to solve the problem of the instability of the quartic coupling, the Landau pole of the $\U(1)_\rmY$ gauge coupling has proved to be a stumbling block. 

Ought the SM to be asymptotically safe?  For all practical purposes the  breakdown of the perturbative regime represented by the presence of the Landau pole in  the $\U(1)_\rmY$ gauge coupling can be ignored for it takes place at energies  well beyond the Planck scale. Be that how it may, the taming of the $\U(1)_\rmY$ Landau pole becomes essential  if we  take the UV behavior  as our guidance in searching for a completion of the SM.
 
A research program based on the safe  UV completion of the SM has been actively pursued in recent years thanks to the progress that has been made in gauge theories with a large number of vector-like fermions and gauge bosons---for which it is possible to state rigorous results~\cite{Litim:2014uca,Bond:2017lnq} in the Veneziano limit. These findings have encouraged the investigation of the extension with vector-like fermions of models containing at least some of the features of the SM~\cite{Holdom:2014hla,Bond:2017wut,Mann:2017wzh,Pelaggi:2017abg} and, more recently, the SM itself~\cite{Barducci:2018ysr,Hiller:2020fbu}. All the same, impressive as these results are, it is fair to say that the Landau pole of the $\U(1)_\rmY$  gauge coupling has proved to be a stumbling block. It appears that all perturbative stable fixed points of the possible extension of the SM with vector-like fermions only admit a low-energy matching  if  the $\U(1)_\rmY$  gauge coupling vanishes and the theory is trivial in that sector. 
 
 This problem behooves us to look into viable options that allow to circumvent the $\U(1)_\rmY$  triviality problem. 

One possible way out has been recently suggested by studying an asymptotically safe version of QED.
It has been shown in \Citeref{Gies:2020xuh} that, an enlarged theory space---where higher dimensional operators such as a Pauli spin-field coupling are included---opens the possibility for UV-complete realizations of QED due to the presence of interacting fixed points.
Other possible solutions to the triviality problem are obtained by including \textit{ad hoc} gravitational contributions~\cite{Eichhorn:2015kea,Alkofer:2020vtb}---a procedure with its own conceptual difficulties---or venturing into  the non-perturbative regime~\cite{Hiller:2020fbu}. 

In this work we follow the more conservative choice of looking into
a grand unified theory (GUT) extension of the SM~\cite{Georgi:1974sy,Pati:1974yy,PhysRevD.11.566,FRITZSCH1975193,GeorgiHoward:1975}, where the abelian gauge group is merged into a larger nonabelian group for which there is no Landau pole to begin with. The possibility of having an asymptotically safe GUT has been discussed in~\cite{Bajc:2016efj,Molinaro:2018kjz}.
Other embeddings that solve the $\U(1)_\rmY$  problem have been proposed in \Citerefs{Giudice:2014tma,Pelaggi:2015kna}.

We study minimal GUT extensions based on the $\SU(5)$ gauge group. 
The specific model we consider below the GUT scale consists in the SM enlarged by  the addition 
of vector-like fermions---that is, fermions whose right- and left-handed components belong to the same representation of the gauge group and for which  a Dirac mass term can be explicitly written. They enter at the scale of 1 TeV and transform under some specific representations of the SM gauge group. 

The role played by the vector-like fermions is twofold:
below the GUT scale their presence lead to a ``good'' (that is, around 1\% of relative difference) gauge coupling unification and
ensures such unification at a scale which is not constrained by low energy experiments like proton decay; 
above the GUT scale, thanks to their new Yukawa interactions, they are crucial for generating non trivial perturbative UV fixed points for the $\SU(5)$ gauge coupling~\cite{Bond:2016dvk}.

In particular, we consider one of the possible minimal extensions that were classified in \Citeref{Giudice:2004tc}: the \underline{{\LVG} model}.
The corresponding TeV scale vector-like fermions are embedded in proper $\SU(5)$ representations at the GUT scale with multiplicities adjusted such that the UV fixed point of the $\SU(5)$ gauge coupling has a numerical value rather close to that of the unification of the three SM gauge couplings at the GUT scale. 
The gauge couplings run from the electroweak to the GUT scale where they come close to each other and merge into the $\SU(5)$ gauge coupling.
From the GUT scale on, the unified gauge coupling remains at its fixed point.

The other relevant couplings that are present below the GUT scale,
namely the top-quark Yukawa, the vector-like fermion Yukawa couplings and the quartic Higgs coupling,
run through the GUT scale where they merge into their $\SU(5)$ GUT counterpart and  reach their own UV fixed points together with the other couplings of the GUT model (additional vector-like fermions Yukawa couplings and GUT scalar potential couplings).

All couplings, the fixed point values and critical exponents always remain in the perturbative regime---a fact that suggests that the fixed point and the renormalization group flow are  stable.

\section{Toward the GUT scale}
\label{sec:bgutmodel}
The gauge couplings of the SM run toward each other in a manner that is suggestive of a possible unification. Though they come rather close, they do so at a scale of order  $10^{13}$-$10^{14}$ GeV, that is too low for the GUT theory  to be consistent with data on proton lifetime~\cite{Miura:2016krn}. On the other hand, it is known that the  addition of new charged states can modify the running and move the GUT scale to a higher value. Among the possible models, minimal non-supersymmetric extensions were discussed in \Citerefs{PhysRevD.45.R3903,Bin_2001,Choudhury_2002,LI_2004,Morrissey_2004,Giudice:2004tc,Dorsner_2005,Emmanuel_Costa_2005,Shrock_2008,Gogoladze:2010in} and \Citerefs{Derm_ek_2012,Derm_ek_2013,Dor_ner_2014,Xiao_2014,Bhattacherjee_2018,
Schwichtenberg_2019}. On the other hand, supersymmetric extensions can be found in \Citerefs{Arkani_Hamed_2005,BARGER_2007,Barger:2007qb,Calibbi_2009,Donkin_2010,Liu_2013}. Yet the minimal supersymmetric  GUT model with squark masses $m_{\tilde f}\lesssim 2$ TeV is excluded if one combines the limits on proton decay mediated by   the colored Higgs 
\cite{Goto:1999iz} with the constraints obtained by the requirement that the Yukawa coupling do not blow up before Planck scale \cite{Hisano:1992mh,Hisano:1994hb}. 

In this work we focus on non-supersymmetric theories and consider one specific minimal extension of the SM that has been classified in \Citeref{Giudice:2004tc}, namely the {\LVG} model, which has the same low energy field content of the ``split-SUSY'' scenario, as summarized in Tab.~\ref{tab:LVG}.
The label ``\LL'' stands for vector-like fermions $\psi_L$ that transform under the $(1, \mathbf{2}, 1/2)$ representation of the SM gauge group, and have the same quantum numbers of the minimal supersymmetric SM Higgsino.
The labels ``\V'' and  ``\G''  stand for Majorana fermions $\psiV$ and $\psiG$  that transforms, respectively, under the $(1,\mathbf{3},0)$ and $(\mathbf{8},1,0)$ representations of the SM.
The fields $\psiV$ and $\psiG$ are like the wino and gluino fields of the minimal supersymmetric SM.
The multiplicities of these beyond the SM representations are all the same and equal to $\Nf=1$.
These extra matter fields are added at the scale of 1 TeV.

\begin{table}[t!]
\begin{center} 
\vspace{0.2cm}
\bgroup
\def\arraystretch{1.3}
\begin{tabular}{|c|c|c|c|c|}
\hline
{\bf Fields} & $\SU(3)_\rmc$ & $\SU(2)_\rmL$ & $\U(1)_\rmY$ & $\Nf$ \cr
 \hline 
  $\psiL$ &  1 & $\mathbf{2}$ & 1/2 & 1 \cr	
   $\psiV$ &  1 & $\mathbf{3}$ & 0 & 1 \cr
 $\psiG$ &  $\mathbf{8}$ & 1 & 0 & 1 \cr  		
\hline 				
\end{tabular}
\egroup
\caption{\small Quantum numbers and multiplicities of the vector-like fermions in the {\LVG} model.} 
\label{tab:LVG}	
\end{center}
\end{table}
The Lagrangian of the {\LVG} model  is given by
\be
{\Lagr}_{\LVG}={\Lagr}_{\rm SM}
+\psibarL\I \slashed D \psiL
-\ML\psibarL\psiL
+\Tr\,\psibarV\I \slashed D \psiV
-\MV\Tr\,\psibarV\psiV
+\Tr\,\psibarG\I \slashed D \psiG
-\MG\Tr\,\psibarG\psiG
-\yV \psibarL \psiV H + {\rm h.c.}
\label{LVGlag}
\ee
where $\psiV=\psiV^i\, T_i$,
with $T_i$ the generators of $\SU(2)_\rmL$ such that $\Tr \, (T_i T_j)=1/2 \, \delta_{ij}$,
and $\psiG=\psiG^a\,\lambda_a$,
with $\lambda_a$ the generators of $\SU(3)_\rmc$ normalized again such that $\Tr\, (\lambda_a \lambda_b)=1/2\,\delta_{ab}$.
The indices $(i,j=1,2,3)$ and $(a,b=1,2,\dots,8)$ belong to the adjoint representations of $\SU(2)_\rmL$ and $\SU(3)_\rmc$ respectively.
Let us notice that the SM Higgs doublet $H$ can couple to the vector-like fermions via a new Yukawa interaction whose coupling is $\yV$.

The SM Lagrangian ${\Lagr}_{\rm SM}$ in \Eqref{LVGlag}  must be thought as written following the convention used in \Citeref{Barducci:2018ysr}, in particular the Higgs quartic interactions is parametrized as $-\lambda\, H^\dagger H/2$.	
We did not include Yukawa interactions that mix SM fermions and vector-like fermions (this can be achieved by imposing  a $\mathbb{Z}_2$/parity-type symmetry, under which SM fermions are even while vector-like fermions are odd).

\subsection{Renormalization group flow}
\label{sec:RG-for-LVG}
In this section we  study the renormalization group flow for the couplings of the \LVG\ model, in what is known as the  \textsc{211-scheme} approximation, where the gauge coupling $\beta$-functions are computed at two-loop order while the Yukawa and scalar couplings are computed at one-loop order.
This scheme is a compromise between the  more formally consistent  \textsc{321-scheme}---a scheme  with $\beta$-functions  at the three-loop order in the gauge coupling, two-loop order in the Yukawa couplings and one-loop order scalar couplings---and the computational manageability  of the $\beta$-functions. The simpler and consistent  \textsc{210-scheme}  would not allow us to study the renormalization of the scalar potential. We trust that higher-loop corrections do not significantly change our results since we always work well within the perturbative regime.

We compute the $\beta$-functions in the $\overline{{\rm MS}}$ renormalization scheme and consider only the gauge, top-Yukawa, Higgs scalar quartic and vector-like fermion Yukawa couplings.
In the rest of this paper we will neglect all other Yukawa couplings in the SM as they are small compared to that of the top quark.
Let us define the following rescaled couplings $\alpha$'s:
\begin{align}
	\label{alphasbgut}
	&\alpha_i=\frac{g_i^2}{(4\pi)^2}\,,&
	&\alphatm =\frac{\yt^2}{(4\pi)^2}\,,&
	&\alpha_{\lambda}=\frac{\lambda}{(4\pi)^2}\,,&
	&\alphaV =\frac{\yV^2}{(4\pi)^2}\,,
\end{align}
where $g_1, g_2, g_3$, $\yt$ and $\lambda$ are the SM couplings and $\yV$ is the vector-like fermion Yukawa coupling. This definitions are convenient in expressing the $\beta$-functions as polynomials with rational coefficients. Note that the definition of $\alpha_i$ for the gauge couplings is different from the usual one by an additional factor of $(4 \pi)$ in the denominator. 
The $\beta$-functions of the \LVG\ model read
\begin{align}
	\label{b1}
	\de_t \alpha_1&= \beta_1^{\rm SM,NLO}+\left(\frac{4}{3}+\alpha_1+3\alpha_2-6\alphaV\right) \alpha_1^2\,,\\
	\label{b2}
	\de_t \alpha_2&= \beta_2^{\rm SM,NLO}+\left(4+\alpha_1+59\alpha_2-22\alpha_V\right) \alpha_2^2\,,\\
	\label{b3}
	\de_t \alpha_3&= \beta_3^{\rm SM,NLO}+\left(4+96\alpha_3\right)\alpha_3^2\,,\\
	\label{bt}
	\de_t \alphat&= \beta_{\rm t}^{\rm SM,LO}+12\alphaV \alphat\,,\\
	\label{bv}
	\de_t \alphaV&= \left(15\alphaV+6\alphat-\frac{3}{2}\alpha_1-\frac{33}{2}\alpha_2\right)\alphaV\,,\\
	\label{bl}
	\de_t \alpha_\lambda &= \beta_\lambda^{\rm SM,LO}+24\alphaV \alpha_\lambda - 48 \alphaV^2\,,
\end{align}
where $\beta_i^{\rm SM,NLO}$, $\beta_{\rm t}^{\rm SM,LO}$ and $\beta_\lambda^{\rm SM,LO}$  are the SM $\beta$-functions given in \Appref{app:smbeta}. 
The new terms pertaining to physics beyond the SM  are explicitly shown: the vector-like fermion contributions to the gauge couplings  are computed using the formulas in \Appref{app:vlfbeta}.
The contributions of the vector-like fermion Yukawa coupling to the gauge, top-Yukawa and Higgs quartic couplings as well as the $\beta$-function of $\alpha_V$ itself are computed using the results of \Citeref{XiaoLuo:2003}.
An explanation of the latter contributions is provided in \Appref{app:vlfbeta}.

\begin{table}[t!]
	\begin{center}
		\bgroup
		\def\arraystretch{1.3}
		\begin{tabular}{|c|c|c|c|c|}
			\cline{1-5} 
			$\mathbf{\alpha_1}$ &  $\mathbf{\alpha_2}$ & $\mathbf{\alpha_3}$ & $\mathbf{\alphat}$ & $\mathbf{\alpha_\lambda}$\cr
			\hline
			$\; 0.0008091 \; $ & $\; 0.002689\;$  &$\;0.009390\;$ & $\;0.006298\;$ &  $\;0.001634\;$\cr
			\hline   
		\end{tabular}
		\caption{\small Initial conditions at $M_Z=91.19$ GeV for the SM gauge, top-Yukawa and Higgs quartic couplings.}  
		\label{tab:IC}
		\egroup
	\end{center}
\end{table}

\begin{figure}[t!]
\begin{center}
\includegraphics[width=0.85\columnwidth]{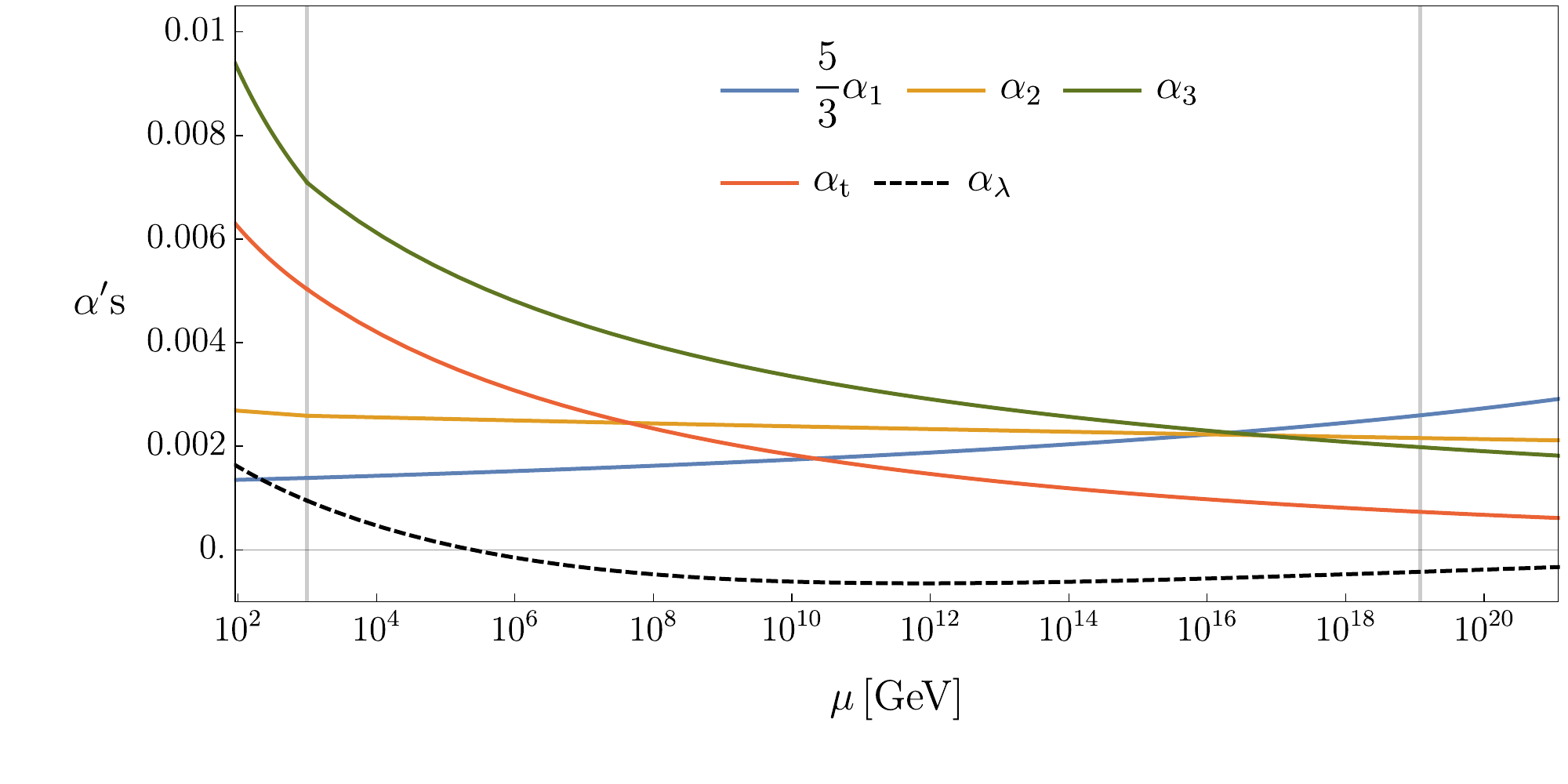}
	\vspace{-5pt}
	\caption{\small Renormalization group flow of the SM couplings $5/3\,\alpha_1$, $\alpha_2$, $\alpha_{3}$, $\alphat$ and $\alpha_\lambda$ for the \LVG\ model.
	The initial conditions are given at the scale $M_Z$ and the vector-like fermions are included at the scale of 1 TeV (gray vertical line on the left).
	The flow is considered up to the Planck scale (gray vertical line on the right).
	}
	\label{fig:BGUTrun}
		\end{center}
\end{figure}

In computing the renormalization group flow, we assume that the vector-like fermions have all the same mass $\ML=\MV=\MG=1$ TeV.
The initial conditions for the SM couplings $\alpha_i$, $\alphat$ and $\alpha_\lambda$ are given at the $Z$-boson mass $M_Z=91.19$ GeV.
These values are shown in \Tabref{tab:IC} and are obtained by using the tree level relations between the couplings and the SM input experimental values~\Citeref{Tanabashi:2018oca}.
The renormalization group flow for the couplings of the \LVG\ model, where  $\alphaV$ is set to zero, is shown in \Figref{fig:BGUTrun}. The gray vertical line on the left corresponds to the scale at which the vector-like fermions become dynamical and their presence makes possible to achieve gauge coupling unification at the scale
\begin{align}
	\text{\MGUT}\simeq 2.399\times 10^{16}\text{ GeV}\,,
	\label{eq:scale-GUTlvg}
\end{align}
which is below the Planck scale which is highlighted in \Figref{fig:BGUTrun} by the gray vertical line on the right.
The values of the SM couplings at the GUT scale are
\begin{align}
	\label{irtarget}
	&\alpha\left(\text{\MGUT}\right)\simeq 0.002247\,,&
	&\alphat\left(\text{\MGUT}\right)\simeq 0.0009388\,,&
	\alpha_\lambda\left(\text{\MGUT}\right)\simeq - 0.0005420 \,.
\end{align}

These values will represent the infrared (IR) target which has to be connected with the ultraviolet (UV) behavior of the model defined at the GUT scale; behavior controlled by the existence of interacting fixed points.
The procedure that we use to connect, whenever possible, a UV fixed point with the IR target of \Eqref{irtarget} will be exhaustively explained in \Secref{sec:RG-flow-&-matching}.
As already mentioned, at the GUT scale, the \LVG\ model becomes embedded into an $\SU(5)$ gauge theory, such that the SM gauge couplings merge into the $\SU(5)$ gauge coupling and the top-Yukawa coupling as well as the Higgs quartic coupling merge into their $\SU(5)$ GUT counterparts.
The latter couplings have to match the values in \Eqref{irtarget} while flowing down from (at least) one fixed point of the $\SU(5)$ GUT theory itself.

\subsection{Higgs potential stability}
The problem of the Higgs potential instability is already present in the SM, and we do not provide a solution to this issue in our model with vector-like fermions.
Indeed, the renormalization group flow, as plotted in \Figref{fig:BGUTrun}, shows that the quartic Higgs coupling $\alpha_\lambda$ becomes negative at around $10^5$ TeV (in the present \textsc{211-scheme}), signaling a vacuum instability or metastability.

The presence of vector-like fermions mitigates this problem as they make $\alpha_\lambda$ to turn negative at higher energies compared to the SM case.
This is because $\alphat$ runs faster towards zero with respect to the SM case, which is, in turn, due to the fact that $\alpha_2$ and $\alpha_3$ run slower than in the SM.  
Yet, considering the field content of our model, we have that the variation of the instability scale with respect to the SM case is minimal (about  26\%).
In principle, one can think of preventing the quartic Higgs coupling from becoming negative, and keeping good gauge coupling unification below the Planck scale, by adding a large number of vector-like fermions in different SM representations, as shown in \Citeref{Kowalska:2019qxm}.
This choice is however not minimal and its study is beyond the scope of this paper.

In addition, it is also known that the metastability scale in the SM moves to higher values when one considers two-loop and three-loop $\beta$-function for $\alpha_\lambda$ ($10^{7}$ GeV and $10^9$ GeV respectively), as shown in \Citerefs{EliasMiro:2011aa,Degrassi:2012ry}.
In the end, the problem still persists and one  either accepts  a metastable potential, as in the SM, or one adds additional fields (for example vector-like fermions or also a scalar singlet) so as to prevent the Higgs quartic coupling to become negative before the Planck scale.

The present discussion about the scalar potential stability is largely accepted within the standard lore of perturbation theory, which intimately associates the instability scale to the existence of a lower bound for the Higgs mass.
However, the presence of such an instability scale becomes questionable as soon as one adopts different approaches.
For example, lattice simulations have shown that the lower bound for the Higgs mass merely arises from consistency conditions imposed on the bare action and no reference to a low-energy stability issue has to be made~\cite{Krive:1976sg,Holland:2003jr,Holland:2004sd,Gerhold:2007yb,Gerhold:2007gx,Gerhold:2009ub,Gerhold:2010bh,Bulava:2012rb}.
The same point of view got also substantiated by functional methods, showing that the conventional lower Higgs-mass bound can even be relaxed given an appropriate consistent definition of the bare action~\cite{Gies:2013fua,Gies:2014xha,Eichhorn:2015kea,Eichhorn:2014qka,Jakovac:2015kka,Gies:2017zwf,Sondenheimer:2017jin,Gies:2017ajd,Reichert:2017puo}.

\section{Above the GUT scale}
\label{sec:gutmodels}
Aiming at an asymptotically safe scenario, we now focus our analysis beyond the GUT scale, where we assume that an underlying $\SU(5)$ symmetry is restored. Consequently,
the vector-like fermions of the \LVG\ model discussed in the previous \Secref{sec:bgutmodel} are embedded into proper multiples of $\SU(5)$, just like it is the case for the SM fields.
This unification group choice seems to be the most natural since $\SU(5)$ not only can play the role of a self-contained unified gauge symmetry~\cite{Georgi:1974sy}, but also shows up in breaking chains of larger GUT groups.
The  Lagrangian of the $\SU(5)$ SM GUT theory is given by~\cite{Ross:1985ai}
\begin{align}
\label{gutlag}
	\Lagr_\SM^\mathrm{GUT}&= - \frac{1}{4}F^a_{\mu\nu}F_a^{\mu\nu}
	+\overline{\chi}_{\bar{5}}\I \slashed{D}\chi_{\bar{5}}+\frac{1}{2}\Tr\left(\overline{\chi}_{10}\I\slashed{D}\chi_{10}\right)
	+\Tr\left[ (D_\mu\Sigma)^\dagger (D^\mu\Sigma)\right] +(D_\mu \Phi)^\dagger (D^\mu \Phi)-V(\Phi,\Sigma)
	\nonumber\\&\quad
	-y_\rmb \overline \chi^c_{\bar{5}} \chi_{10}\Phi^*+
	\frac{\ytM}{8}\varepsilon_{5}\overline \chi_{10}^c\chi_{10}\Phi+\mathrm{h.c.}\,,
\end{align}
where $F_{\mu\nu}^a$ is the field strength of the $\SU(5)$ gauge fields $A_\mu^a$ ($a=1,..,24$), which include the SM gluons, the electroweak gauge bosons and the heavy GUT gauge bosons.

The right-handed down quarks and the left-handed lepton doublets are embedded into the left-handed field $\chi_{\bar 5}$ transforming as an anti-fundamental $\mathbf{\bar 5}$ representation of $\SU(5)$, while the left-handed quark doublets, right-handed up quarks and leptons are embedded into the left-handed field $\chi_{10}$ transforming as an anti-symmetric $\mathbf{10}$ representation of $\SU(5)$.
The charge conjugation of a fermionic field is expressed by a superscript $c$, for example $\chi^c_{\bar{5}}=\mathcal{C}(\overline \chi_{\bar{5}})^\rmT$ where $\mathcal{C}$ is the charge conjugation operator.

The Higgs field is embedded into $\Phi$, transforming as a fundamental $\mathbf{5}$ representation of $\SU(5)$.
Even though the fermionic matter content of the SM can be fitted entirely into the $\mathbf{\bar{5}}$ and $\mathbf{10}$ representations of $\SU(5)$, the scalar sector is extended by an adjoint scalar field $\Sigma$, in the $\mathbf{24}$ representation, which is needed in order to break the $\SU(5)$ into the SM gauge group.

The Lagrangian in \Eqref{gutlag} includes only the top- and bottom-Yukawa couplings,
which are the most relevant ones.
We set to zero the Yukawa couplings of the first and second generations.
The fermionic field $\chi_{10}$ is a $5\times 5$ anti-symmetric matrix such that the corresponding Yukawa interaction is constructed by mean of the Levi-Civita tensor in five dimensions, i.e., $\varepsilon_{5}$.
In terms of components the Yukawa interaction term is $\varepsilon_{5}\overline \chi_{10}^c\chi_{10}\Phi = \varepsilon_{ijklm}\overline \chi_{10}^{c\,\,ij}\chi_{10}^{kl}\Phi^m$.

The quartic terms of the scalar potential $V(\Phi,\Sigma)$ in \Eqref{gutlag} read~\cite{Ross:1985ai}:
\begin{align}
	V^{\rm quartic}(\Phi,\Sigma)&=\frac{\lambda_\Sigma}{2}\,\Tr\left(\Sigma^4\right)
	+\frac{\lambda_\Sigma'}{2}\,\left[\Tr\left(\Sigma^2\right)\right]^2
	+\frac{\lambda_H}{2}\,\left(\Phi^\dagger\Phi\right)^2
	+2\lambda_{H\Sigma}\Phi^\dagger\Phi \Tr\left(\Sigma^2\right)
	+2\lambda_{H\Sigma}'\Phi^\dagger\Sigma^2\Phi\,,
\label{eq:scalar-potential-SU(5)}
\end{align}
where $\lambda_X=(\lambda_\Sigma$, $\lambda_\Sigma'$, $\lambda_H$, $\lambda_{H\Sigma})$ are the quartic scalar couplings of the GUT model.

At the unification scale, the vector-like fermions of the \LVG\ model in \Tabref{tab:LVG} are embedded into proper multiples of $\SU(5)$.
We assume that the field $\psiL$ gets embedded into vector-like fermions $\Psi_5$, with multiplicity $N_5\geq 1$, transforming under the fundamental representation of $\SU(5)$;
while $\psiV$ and $\psiG$ get embedded into Majorana fermions $\Psi_{24}$, with multiplicity $N_{24}\geq 1$, transforming under the adjoint representation of $\SU(5)$.~\footnote{
The fundamental and adjoint representations of $\SU(5)$ can be decomposed under the SM gauge group as follows (see for example \Citeref{SLANSKY19811}): $\mathbf{5}=(1,\mathbf{2},1/2)\,\oplus\,(\mathbf{3},1,-1/3)$ and $	\mathbf{24}=(1,1,0)\,\oplus\,(1,\mathbf{3},0)\,\oplus\,(\mathbf{8},1,0)\,\oplus\,(\mathbf{3},\mathbf{2},-5/6)\,\oplus\,(\mathbf{\bar{3}},\mathbf{2},5/6)$.}

\begin{table}[t!]
\begin{center} 
\vspace{0.2cm}
\bgroup
\def\arraystretch{1.3}
\begin{tabular}{|c|c|c|}
\hline
{\bf Fields} & $\SU(5)$ & $\Nf$ \cr
 \hline 
 $\Psi_5$ &  $\mathbf{5}$ & $N_5$\cr
 $\Psi_{24}$ &  $\mathbf{24}$ & $N_{24}$\cr		
\hline					
\end{tabular}
\egroup
\caption{\small Quantum numbers and multiplicities of the vector-like fermions in the \LVG\ GUT model.} 
\label{tab:LVGGUT}
\end{center}
\end{table}

Therefore, the Lagrangian of the \LVG\ GUT model is
\begin{align}
	\Lagr^\mathrm{GUT}_{\LVG}&=\Lagr_\SM^\mathrm{GUT}
	+\overline{\Psi}^I_{5}\I \slashed D \Psi_{5}^I
	- M_5\overline{\Psi}^I_{5}\Psi_{5}^I
	+ \Tr\left(\overline{\Psi}_{24}^J	\I \slashed D \Psi_{24}^J\right)
	-M_{24} \Tr\left(\overline{\Psi}_{24}^J	\Psi_{24}^J\right)
	\nonumber\\&\quad
	- y_{5\Sigma}\overline{\VLF}_{5}^I\Sigma\VLF_{5}^I
	- y_{24\Sigma}\Tr\left(\Sigma\overline{\VLF}_{24}^J\VLF_{24}^J\right)
	 -y_\nu \sum_{I,J}\left(\overline{\Psi}_5^I \Psi_{24}^J \Phi + {\rm h.c.}\right)\,,
	\label{eq:lagrLVGGUT}
\end{align}
where $I$ and $J$ run over the flavor numbers, i.e.,  $I=1,\dots,N_5$ and $J=1,\dots,N_{24}$.
In the \LVG\ GUT Lagrangian of the latter equation we have introduced three Yukawa couplings involving the vector-like fermions:
$y_{5\Sigma}$, $y_{24\Sigma}$ and $y_{\nu}$.
The first two terms in the second line respect an $\U(N_5)\times\OO(N_{24})$ flavor symmetry which is explicitly broken by $y_\nu$ interactions.
In principle, additional gauge invariant Yukawa terms are possible, however, we restrict the number of such interactions to just three by imposing a $\mathbb{Z}_2$/parity-type symmetry, under which the SM fermions are even while the vector-like fermions are odd.

\subsection{The $\beta$-functions}
\label{sec:LVGbetas}
In this section we compute the  $\beta$-functions of the \LVG\ GUT model given in \Eqref{eq:lagrLVGGUT}.
We adopt again the $\overline{{\rm MS}}$ renormalization scheme and the \textsc{211-scheme} regarding the loop orders.
Let us first define the following rescaled couplings $\alpha$'s
\begin{align}
\label{eq:def-alphas-AGUT}
	&\alphaM  = \frac{g_5}{4\pi}\,,&
	&\alpha_{X}=\frac{\lambda_X}{(4\pi)^2}\,,& &\alpha_z =\frac{y_z}{4\pi}\,,
\end{align}
where $g_5$ is the $\SU(5)$ gauge coupling, $\lambda_X$ denotes the couplings in the scalar potential $V^{\rm quartic}(\Phi,\Sigma)$ and  $y_z=(\ytM,\,y_{5\Sigma},\, y_{24\Sigma},\, y_{\nu})$ represents all Yukawa couplings.
Note that, differently from the previous \Secref{sec:RG-for-LVG}, here we define all couplings $\alpha$'s as linear with respect to the couplings in Eqs.~(\ref{gutlag}--\ref{eq:lagrLVGGUT}).
These definitions allow us to write the $\beta$-functions as polynomials in all the $\alpha$'s, as it is made  clear  by the equations below.
In the \textsc{211-scheme}, the gauge-Yukawa subsystem is closed and its $\beta$-functions are
\begingroup
\allowdisplaybreaks
\begin{align}
\label{bggut}
\de_t\alphaM &= - \frac{40-2N_5-10N_{24}}{3}\,\alphaM^3
-\frac{1184-322 N_5-2000N_{24}}{15}\,\alphaM^5\nn\\
&\quad-\bigg(\frac{9}{2}\alphatM^2 +\frac{12}{5}N_5\alpha_{5\Sigma}^2+\frac{21}{4}N_{24}\alpha_{24\Sigma}^2 
+\frac{37}{5}N_5N_{24}\alpha_\nu^2 \bigg)\alphaM^3\,,\\
\label{btgut}
\de_t\alphatM&=\left(6\alphatM^2+\frac{24}{5}N_5N_{24}\alpha_\nu^2-\frac{108}{5}\alphaM^2\right)\alphatM\,,
\\
\label{b5gut}
\de_t\alpha_{5\Sigma}&=\Bigg(\frac{11+5N_5}{5}\alpha_{5\Sigma}^2+\frac{21}{20}N_{24}\alpha_{24\Sigma}^2+\frac{12}{5}N_{24}\alpha_\nu^2
-\frac{72}{5} \alphaM^2\Bigg)\alpha_{5\Sigma}+\frac{21}{10}N_{24}\alpha_\nu^2\alpha_{24\Sigma}\,,\\
\label{b24gut}
\de_t\alpha_{24\Sigma}&=\Bigg(\frac{34+21N_{24}}{20}\alpha_{24\Sigma}^2+N_{5}\alpha_{5\Sigma}^2+N_5\alpha_\nu^2
- 30\alphaM^2\Bigg)\alpha_{24\Sigma}+2N_5\alpha_\nu^2\alpha_{5\Sigma}\,,\\
\label{bngut}
\de_t\alpha_\nu &= \Bigg( \frac{63}{10}N_{5}N_{24}\alpha_\nu^2+3\alphatM^2+\frac{21}{40}\alpha_{24\Sigma}^2
+\frac{6}{5}\alpha_{5\Sigma}^2+\frac{21}{10}\alpha_{24\Sigma}\alpha_{5\Sigma}
-\frac{111}{5}\alphaM^2
\Bigg)\alpha_\nu\,.
\end{align}
\endgroup
The choice of defining all the $\alpha$'s as linear with respect to the original couplings becomes clear from the last term in \Eqref{b5gut} and \Eqref{b24gut}.
These two contributions arise from the mixed Yukawa interaction among $\Psi_5$ and $\Psi_{24}$ (last term in \Eqref{eq:lagrLVGGUT}).
In case we were defining $\alpha_{5\Sigma}$ and $\alpha_{24\Sigma}$ as quadratic in $y_{5\Sigma}$ and $y_{24\Sigma}$,
the last term in \Eqref{b5gut} and \Eqref{b24gut} would have involved square root of $\alpha_{5\Sigma}$ and $\alpha_{24\Sigma}$, rendering some of the fixed points in the following \Tabref{tab:lvgfps-all} inaccessible.
For homogeneity we kept linear also the other Yukawa couplings as well as the gauge coupling.
The  $\beta$-functions of the scalar couplings are 
\begingroup
\allowdisplaybreaks
\begin{align}
\label{eq:beta-lambda-Sigma}
\de_t\alpha_{\Sigma}&=\bar \beta_\Sigma+ \left( 4 N_5 \alpha_{5\Sigma}^2+\frac{21}{5}N_{24}\alpha_{24\Sigma}^2\right)\alpha_{\Sigma} - 4N_5\alpha_{5\Sigma}^4+\frac{7}{20}N_{24}\alpha_{24\Sigma}^4\,,\\
\label{eq:beta-lambda-Sigmap}
\de_t\alpha_{\Sigma}'&=\bar \beta _{\Sigma}'+ \left( 4 N_5 \alpha_{5\Sigma}^2+\frac{21}{5}N_{24}\alpha_{24\Sigma}^2\right)\alpha_{\Sigma}'
 -\frac{91}{100}N_{24}\alpha_{24\Sigma}^4\,,\\
\label{eq:beta-lambda-H}
\de_t\alpha_{H}&=\bar \beta_H+\left(12 \alphatM^2+\frac{96}{5}N_5N_{24}\alpha_\nu^2\right)\alpha_H - 12 \alphatM^4-\frac{264}{25}N_5^2N_{24}^2\alpha_\nu^4\,,
\\
\label{eq:beta-lambda-HSigma}
\de_t\alpha_{H\Sigma}&=\bar \beta_{H\Sigma}
+\left( 6\alphatM^2+ 2N_{5}\,\alpha_{5\Sigma}^2+\frac{21}{10}N_{24}\alpha_{24\Sigma}^2+\frac{48}{5}N_5N_{24}\alpha_\nu^2\right)\alpha_{H\Sigma}\nn\\
&\quad -N_5N_{24}\left(2\alpha_{5\Sigma}^2+\frac{29}{50}\alpha_{24\Sigma}^2+\alpha_{5\Sigma}\alpha_{24\Sigma}\right)\alpha_\nu^2\,,
\\
\label{eq:beta-lambda-HSigmap}
\de_t\alpha_{H\Sigma}'&=\bar\beta_{H\Sigma}'
+\left( 6 \alphatM^2
+2N_{5}\,\alpha_{5\Sigma}^2
+\frac{21}{10}N_{24}\alpha_{24\Sigma}^2+\frac{48}{5}N_5N_{24}\alpha_\nu^2\right)\alpha_{H\Sigma}'\\
&\quad -N_5N_{24}\left(-\frac{2}{5}\alpha_{5\Sigma}^2+\frac{13}{10}\alpha_{24\Sigma}^2-\frac{4}{5}\alpha_{5\Sigma}\alpha_{24\Sigma}\right)\alpha_\nu^2\,,
\end{align}
\endgroup
where the pure gauge e scalar contributions $\bar \beta_\Sigma$, $\bar \beta_\Sigma'$, $\bar \beta_H$, $\bar \beta_{H\Sigma}$ and $\bar \beta_{H\Sigma}'$ are given in \Appref{app:su5betas}.
In particular, in \Appref{app:su5betas-scalar-sector} we also give a thorough explanation on how we have derived the extra contributions to $\de_t\alpha_X$ arising from the presence of the vector-like fermion Yukawa couplings.
In \Appref{app:su5betas-gaugeYukawa-sector} we also give a derivation of the $\beta$-functions for the gauge-Yukawa subsystem.

\subsection{Multiplet mass splitting}
\label{sec:splitting}

At the GUT scale, the vector-like fermion fields $\psiL$, $\psiV$ and $\psiG$ are embedded into $\Psi_5$ and $\Psi_{24}$ multiplets, as described in the previous section.
These $\SU(5)$ multiplets contain extra fields and therefore one has to devise a mechanism that, after $\SU(5)$ breaking, keeps these extra components at the GUT scale and allows the $\psiL$, $\psiV$ and $\psiG$ fields to acquire a mass of order 1 TeV.
For instance, let us consider the $\psiL$ field and its embedding into $\Psi_5=\psiL\oplus\psi_{\bf 3}$, where $\psi_{\bf 3}$ is the partner that transforms as an $\SU(3)$ triplet.
After $\SU(5)$ breaking due to a nonzero vacuum expectation value for the adjoint scalar field, namely $\langle \Sigma \rangle = f\cdot{\rm diag}(2,2,2,-3,-3)$, the contributions to the mass term of $\Psi_5$ in \Eqref{eq:lagrLVGGUT} can be written as follows
\be \label{p5mass}
-(M_5+2f y_{5\Sigma} )\overline \psi_{\bf 3} \psi_{\bf 3} -(M_5-3f y_{5\Sigma} )\overline \psi_{L} \psiL\,.
\ee 

Since $M_5\sim f\sim M_{\rm GUT}$, the doublet-triplet splitting can be achieved at tree level by assuming a cancellation between the two terms contributing to the mass term of $\psiL$ in \Eqref{p5mass}, with a tuning of one part in $10^{ 13}$.
A similar mechanism has to be realized also in the case of the $\psiV$ and $\psiG$ embedding into $\Psi_{24}$.
Furthermore, the case of the $\psiL$ embedding is analogous to the doublet-triplet splitting of the Higgs multiplet in minimal $\SU(5)$ GUT models.
The simplest solution to the Higgs doublet-triplet splitting problem is similar to the one in \Eqref{p5mass} and requires a $10^{-14}$ tuning, but it is also possible to devise alternative mechanisms that alleviate this tuning, like the sliding singlet mechanism~\cite{Witten:1981kv,Nanopoulos:1982wk,Dimopoulos:1982af,Ibanez:1982fr}.

\section{Fixed points }
\label{sec:fixed-point}

Consider a theory characterized by a set of dimensionless couplings $\alpha_{\rm i}$. The renormalization group flow is completely determined by their $\beta$-functions
\be
 \beta_{\rm i} (\alpha_{\rm j})\equiv \frac{\de \alpha_{\rm i}}{\de t}\,,
 \ee
where $t=\log(\mu/\mu_0)$ is the logarithm of the sliding scale of the quantum theory. A fixed point of the theory $\alpha^{*}_{\rm j}$ is defined by the vanishing of the $\beta$-functions of all couplings
\be
\beta_{\rm i} (\alpha_{\rm j}^*)=0\,.
\ee
When the  couplings $\alpha_{\rm j}$ assume the values $\alpha^{*}_{\rm j}$, the renormalization of the quantum theory stops. In general, a given fixed point can be reached either in the UV or in the infrared (IR) limit, depending on the direction of the approaching trajectory.
Notice that, in the common lore, the distinction between UV and IR fixed points is only meaningful when there is a single coupling in the theory.
In the case of more couplings, this distinction becomes unambiguous only if, given two fixed points, it exists an RG trajectory connecting the two of them.

The $\beta$-function of a single coupling is
independent of the gauge choice in dimensional regularization. 
It is regularization scheme-independent up to NLO. If there are several  couplings running together, their $\beta$-functions depend on the scheme already at the NLO~\cite{McKeon:2017mjq}. There is therefore a degree of ambiguity in the position of the fixed points because their position could be moved by changing the scheme. We assume that these changes are small if the fixed point is found within the perturbative regime.
One should however bear in mind this problem of scheme dependence in all the discussions to follow.

Once we have a candidate fixed point, we can study the flow in its immediate neighborhood. We move away from the fixed point and study what happens when we shift the couplings by a small amount $\delta\alpha_{\rm i}\equiv \alpha_{\rm i} - \alpha^{*}_{\rm i}$. To this end, we linearize the $\beta$-functions in the vicinity of the fixed point as
\be
\label{Stability}
\frac{\de}{\de  t}\delta\alpha_{\rm i}=M_{\rm ij}\,\delta\alpha_{\rm j}
\ee
and ignore ${\cal{O}}\left(\delta\alpha^2\right)$ terms. The quantity
\be
M_{\rm ij}\equiv \frac{\de\beta_{\rm i}}{\de \alpha_{\rm j}}\bigg|_{\alpha_{\rm i}^*}	
\ee
 is referred to as the {\it stability matrix}. Next, we can diagonalize the linear system by means of a similarity transformation 
\begin{equation}
    (S^{-1})_{\rm ij}M_{\rm jl}S_{\rm ln}=\delta_{\rm in}\theta_{\rm n} \, ,
\end{equation}
where the eigenvalues $\theta_{\rm n}$ are also known as critical exponents (see equation below). Defining $z_{\rm i} =(S^{-1})_{\rm ij}\, \delta\alpha_{\rm j}$, we have that the $\beta$-functions and their solutions can be written in the following simplified form  
\begin{equation}
    \frac{\de z_{\rm i}}{\de t}=\theta_{\rm i}z_{\rm i}\, \quad \implies \quad \ z_{\rm i}(t)=c_{\rm i}\, \E^{\theta_{\rm i}t}=c_{\rm i}\left( \frac{\mu}{\mu_{0}} \right)^{\theta_{\rm i}}.
\end{equation}
From the expression of $z_{\rm i}$ as functions of $\mu$, we see that there are different situations depending on the sign of $\theta_{\rm i}$:
\begin{itemize} 
\item For $\theta_{\rm i}>0$, as we increase $\mu$ we move away from the fixed point and $z_{\rm i}$ increases without control; the direction $z_{\rm i}$ is said to be {\it irrelevant}.
\item If $\theta_{\rm i}<0$, as we increase $\mu$ we approach the fixed point; 
 the direction $z_{\rm i}$ is called a {\it relevant} direction.
 \item If $\theta_{\rm i}=0$, we do not know the fate of $z_{\rm i}$ and we have to go beyond the linear order as explained below;
the direction $z_{\rm i}$ is called {\it marginal} in this case.
\end{itemize}
The notion of relevance or irrelevance is independent of the direction of the flow and of the choice of basis.
AS theories correspond to trajectories lying on a critical surface whose tangent space at the fixed point is spanned by the relevant eigenvectors.
The number of relevant directions defines the dimension of the critical surface and corresponds to the number of free parameters which have to be fixed by the experiment.

\begin{figure}[t!]
	\includegraphics[width=0.4\columnwidth]{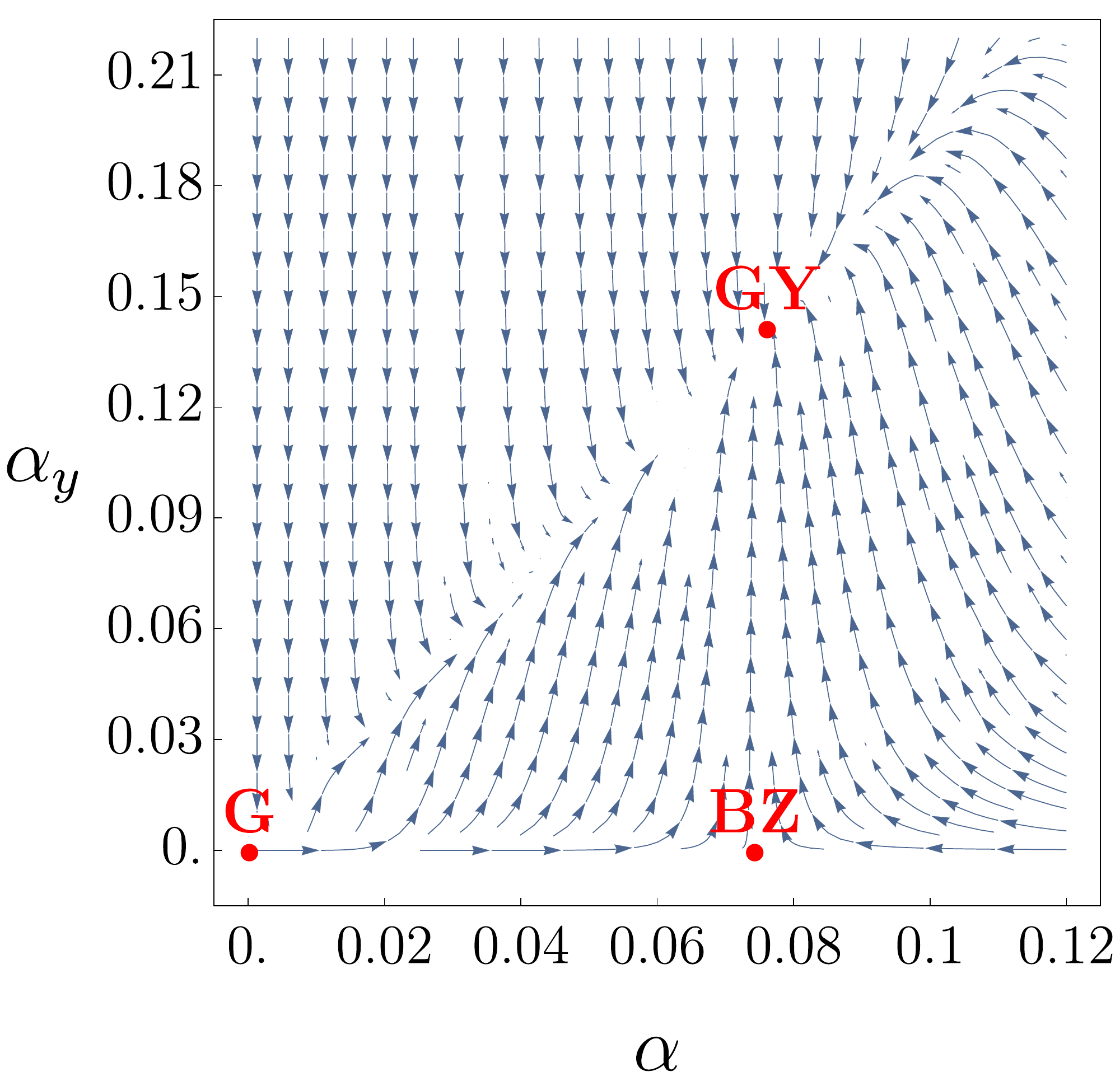}
	\caption{Example of  renormalization group flow in a model with gauge $g$ and Yukawa $y$ couplings.
	The arrows are pointing towards the IR.
	There are three fixed points: the Gau\ss ian (G) for vanishing gauge and Yukawa couplings, the Banks-Zaks (BZ) for vanishing Yukawa coupling and the full interacting Gauge-Yukawa (GY) fixed point.
	}
	\label{fig:stream}
\end{figure}

Gauge-Yukawa models in the  \textsc{211-scheme}, 
which are characterized by the following set of  $\beta$-functions:
\bea
\label{gaugeyuksystem}
\de_t\alpha &=&\biggl(-B +C\alpha -\sum_{\rm i}D_{\rm i}\alpha_{y_{\rm i}}\biggr)\alpha^2\,,\nn\\
\de_t\alpha_{y_{\rm i}} &=&\biggl(\sum_jE_{\rm ij}\alpha_{y_{\rm j}} -F_{\rm i}\alpha\biggr)\alpha_{y_{\rm i}} \, ,
\eea
have been recently studied in \Citerefs{Bond:2016dvk,Bond:2017lnq,Bond:2017wut}. In the latter equation the couplings are $\alpha=g^2/(4\pi)^2$ and $\alpha_{y_{\rm i}}=y_{\rm i}^2/(4\pi)^2$, where $g$ is the gauge coupling and $y_{\rm i}$  are the Yukawa couplings.
The quantities $B$, $E_{\rm ij}$ and $F_{\rm i}$ are the one-loop coefficients, while $C$ and $D_{\rm i}$ are the two-loop ones.
It has been show that, depending on the relative sign and magnitude of the coefficients
$B$, $C$ and $C'=C-\sum_{\rm ij}D_{\rm i}(E^{-1})_{\rm ij}F_{\rm j}$, this system can have three different types of fixed points~\cite{Bond:2016dvk}:
\begin{itemize}
	\item The Gau\ss ian or non-interacting fixed point, where all couplings are zero.
	\item The Banks-Zaks fixed point~\cite{Banks:1981nn} where all the Yukawa couplings vanish.
	\item The Gauge-Yukawa fixed point, where the gauge and at least one Yukawa coupling are different from zero.
\end{itemize}  
As an example, the phase diagram for the case of a gauge coupling $\alpha$ and a single Yukawa coupling $\alpha_y$ with $B>0$ and $C>C'>0$ is shown in \Figref{fig:stream}. In this case all three kind of fixed points are present: the Gauge-Yukawa fixed point has both non vanishing couplings and attracts in the IR trajectories emanating in the UV from both the Gau\ss ian and the Banks-Zaks fixed points. 

In the minimal $\SU(5)$ GUT model, the system of gauge and top-Yukawa $\beta$-functions takes the form of \Eqref{gaugeyuksystem} with $B>0$ and $C<0$: in this case no Gauge-Yukawa fixed point is present and the only viable fixed point is the Gau\ss ian one.
However, the \LVG\ GUT model considered in the previous section has additional vector-like fermions and Yukawa couplings which influence the sign of the coefficients $B$, $C$ and $C'$.
This will allow Gauge-Yukawa fixed points to appear, as we will see in the following.
Actually, the gauge-Yukawa system of our GUT model, described in Eqs.~(\ref{bggut}--\ref{bngut}), is slightly different from the one in \Eqref{gaugeyuksystem} due to non-factorizable contributions of $\alpha_\nu$ to the  $\beta$-functions of $\alpha_{5\Sigma}$ and $\alpha_{24\Sigma}$.
This is indeed the reason why we have considered linear rescaling in the definition of $\alpha$'s in \Eqref{eq:def-alphas-AGUT}.
Nonetheless, the structure of the gauge $\beta$-function remains the same.

In this work we compute the fixed points of the full system, including the gauge, Yukawa and scalar quartic couplings.
In general, there are no conditions on the values of the fixed points and they could take any value.
Yet when we work in perturbation theory, we have to remain within its range of validity.
Therefore, we demand that all the couplings have to be sufficiently small at the fixed point.
In practice this means that going to the next order of the perturbative expansion should not appreciably change the position of the fixed point as well as its other properties.
This implies that the numerical values of the fixed points must satisfy the conditions
\begin{align}
\label{eq:conditions-for-the-FPs}
 &0\leq \alphaM^* < 1\,,&
 &|\alphatM^*|  < 1\,,&
 &|\alpha_z^*|<1\,, &
 &|\alpha_X^*|<1\,.
\end{align}
The complete list of non-trivial fixed points that satisfy these requirements, as function of $N_5$ and  $N_{24}$, is shown in \Tabref{tab:lvgfps-all}.
By inspection of \Tabref{tab:lvgfps-all} we can see that interacting fixed points can be obtained when $N_5=1,2,3,4$ and $N_{24}=3$.
In this case the two-loop term of the gauge $\beta$-function turns out to be comparable with the one-loop term as $C\gg B\gtrsim 0$ and the gauge coupling fixed point turns out to be much smaller than one.
The associated critical exponents, which are shown in \Tabref{tab:lvgfps-all-eigenvalues}, are also much smaller than one and, therefore, we expect the fixed points to be perturbative stable.

\begin{table}[t!]
	\centering
	\includegraphics[width=\columnwidth]{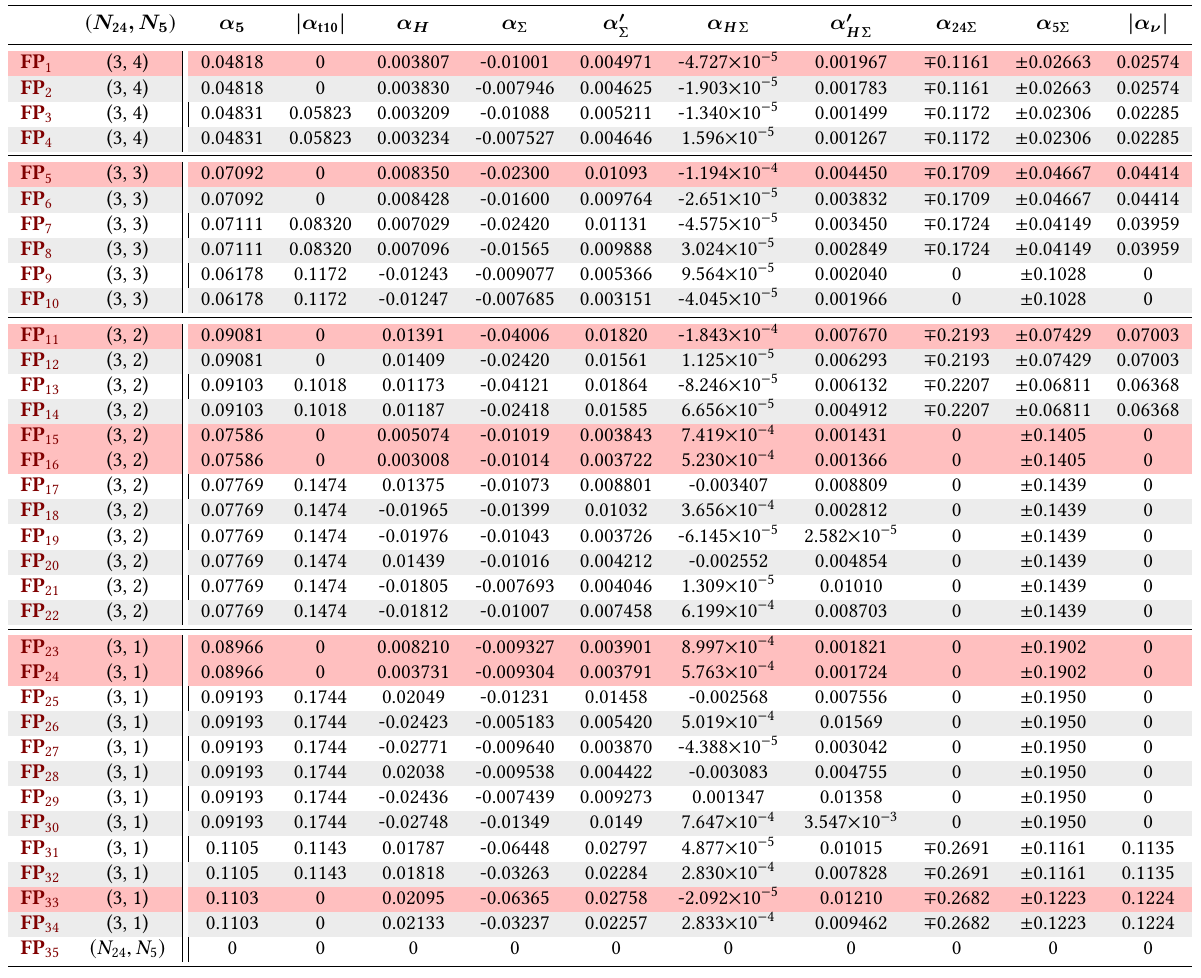}
	\caption{\small
	List of all fixed points for the \LVG\ GUT model which satisfy the conditions in \Eqref{eq:conditions-for-the-FPs}, for various multiplicities of the vector-like fermions $\Psi_{24}$ and $\Psi_5$.
	Those fixed points which possess physical trajectories that can be matched with the \LVG\ model at the GUT scale are highlighted by red rows.
	These fixed points provide therefore an asymptotically safe $\SU(5)$ GUT completion of the SM.
	Each fixed point FP$_n$ has a certain degeneracy (in the sense that physical properties such has the number of relevant/irrelevant directions as well as the critical exponents are identical) due to the fact that the couplings $\alphatM$, $\alpha_{24\Sigma}$, $\alpha_{5\Sigma}$ and $\alpha_{\nu}$ can have both positive or negative sign.
	While $\alphatM$ and $\alpha_{\nu}$ can each be positive or negative independently on the sign of the other coupling, $\alpha_{24\Sigma}$ and $\alpha_{5\Sigma}$ have always opposite signs.
	As an example FP$_1$ incorporates 4 degenerate fixed points while FP$_3$ encodes 8 degenerate fixed points.}	
\label{tab:lvgfps-all}
\end{table}

\begin{table}[t!]
	\centering
	\includegraphics[width=\columnwidth]{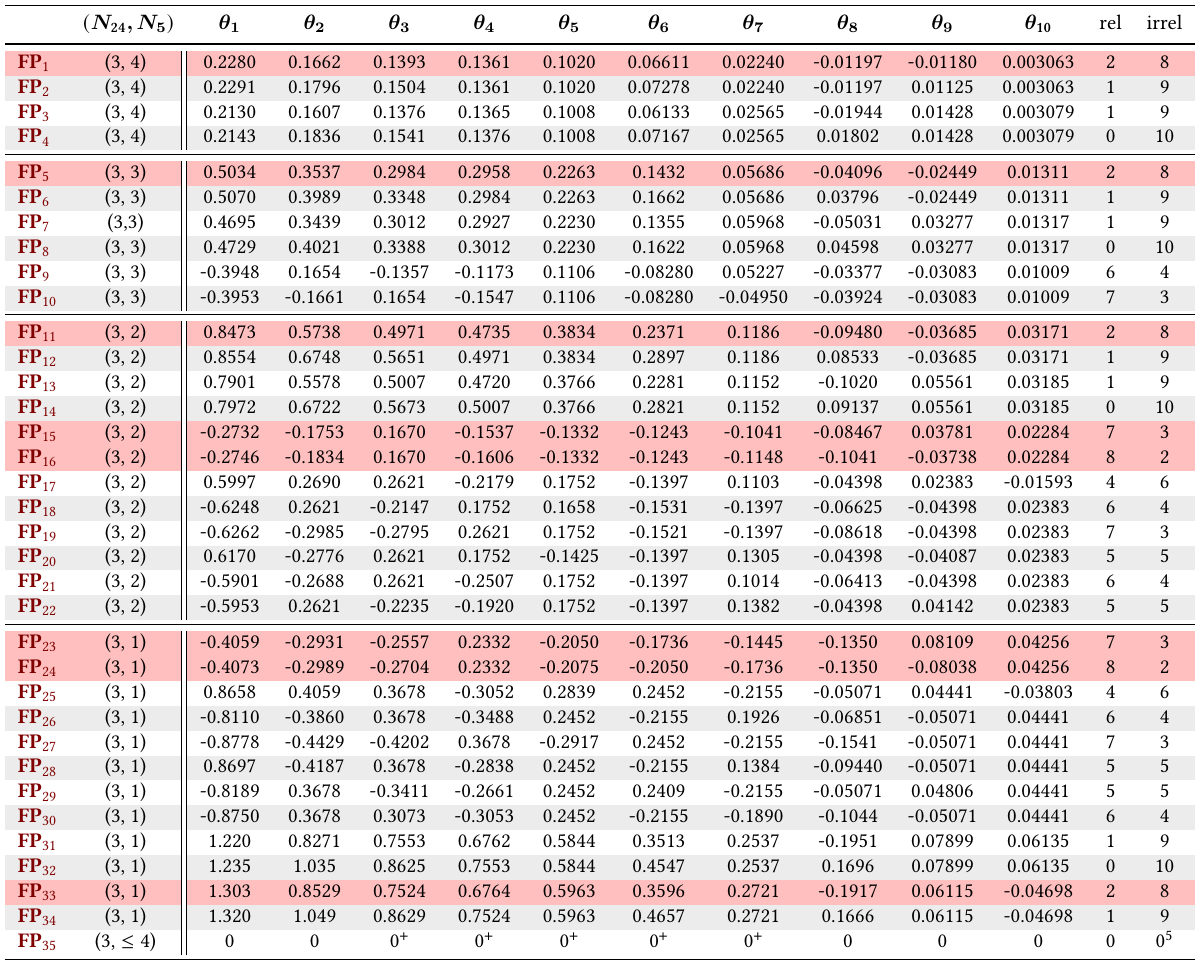}
	\caption{\small	
	List of the eigenvalues for the fixed points shown in \Tabref{tab:lvgfps-all}.
	Negative (positive) eigenvalues correspond to relevant (irrelevant) directions, accordingly to the definitions given in \Secref{sec:fixed-point}.
	As for the \Tabref{tab:lvgfps-all}, the red rows highlight those fixed points which can be matched with the low energy \LVG\ model at the GUT scale.
	For the Gau\ss ian fixed point FP$_{35}$ all couplings are marginals, in particular there are no marginal relevant directions to perturb along which.}
	\label{tab:lvgfps-all-eigenvalues}
\end{table}

\section{Renormalization group flow and matching with the SM}
\label{sec:RG-flow-&-matching}
In this section we investigate the existence of asymptotically safe trajectories that emanate from the UV fixed points presented in \Tabref{tab:lvgfps-all} and are connected to the SM at the IR scale $\mu=M_Z$.
As already mentioned, we assume that the vector-like fermions $\psiL$, $\psiV$ and $\psiG$ -- introduced in \Secref{sec:bgutmodel} --
have a common mass $M_\psi=1$ TeV, while the vector-like fermions $\Psi_5$ and $\Psi_{24}$ -- introduced in \Secref{sec:gutmodels} -- have a common mass at the GUT scale. 
Therefore we need to consider two matching scales
\begin{equation}
	\mu_1=M_\psi=1\text{ TeV}\qquad {\rm and}\qquad
	\mu_2=M_\Psi=\text{\MGUT}\,,
\end{equation}
such that $M_Z<M_\psi\ll M_\Psi$.

We approximate the decoupling of all vector-like fermions by considering them as massless above their corresponding matching scale and as infinitely  massive below.

The running of the various couplings must be matched at the GUT scale, on the interface between the two models described in \Secrefs{sec:bgutmodel}~and~\ref{sec:gutmodels}.
At this interface, a subtlety regarding the gauge couplings should be taken into account;
since their $\beta$-functions have been considered at two-loop order, one-loop matching corrections should consequently accounted for.
The latter corrections are given by the following expression (c.f.~for example \Citerefs{Weinberg:1980wa,Hall:1980kf})
\begin{align}
\label{eq:Hall-matching-eq}
\frac{1}{\alpha_i(\mu_2)} =\frac{1}{\alpha(\mu_2)}- \frac{1}{3} \left[ \Tr   \left(t_{i, \rmV}^2\right)- 21\, \Tr \left( t_{i, \rmV}^2 \log \frac{M_\rmV}{\mu_2}\right)
+  \Tr \left( t_{i, \rmS}^2 \log \frac{M_\rmS}{\mu_2}\right) 
+ 8\, \Tr \left( t_{i, \rmF}^2 \log \frac{M_\rmF}{\mu_2}\right)  \right]\,,
\end{align}
where $M_{\rm X}$ with X=(V, S, F) are the mass matrices for the heavy vector bosons (heavy gauge bosons, ghosts and Goldstone bosons along the broken directions in the adjoint space of $\SU(5)$),
heavy physical scalars and heavy fermions respectively.
The matrices $t_{i,{\rm X}}$ are, instead, the unbroken generators in the representation of the field ${\rm X}$, while the trace is performed over the broken subspace of $\SU(5)$.
In physical terms, the latter \Eqref{eq:Hall-matching-eq} entails that the renormalized gauge couplings $\alpha_i$ 
(after the gauge group $\SU(5)$ is broken down to the SM gauge group)
differ from the gauge coupling $\alpha$ of $\SU(5)$ due to the one-loop diagrams where the quantum fluctuations of the heavy particle are integrated out.
Numerically, the most important contribution comes from the vector states, if they have a mass different from the matching scale $\mu$.
Therefore, we are free to bridge any---reasonably small---mismatch in the running of the gauge couplings by removing the degeneracy in the values of the heavy masses and slightly moving them away from the GUT scale.

\vspace{\baselineskip}

The values of the \LVG\ model couplings $\alpha_i$, $\alphatm$, $\alpha_\lambda$ and $\alpha_V$ at \MGUT\ define the IR target for the \LVG\ GUT model couplings $\alphaM^2$, $\alphatM^2$, $\alpha_H$ and $\alpha_\nu^2$ respectively.
In other words, one has to search for those trajectories emanating from the UV fixed points that hit the values of the IR target at the scale \MGUT, while flowing down from the fixed points.
The other couplings of the \LVG\ GUT model will be consequently determined 
by this matching condition requirement.
Because of the freedom of choosing the mass of the heavy particles $M_{\rm X}$ slightly different from the matching scale $\mu_2$,
for any practical purposes, one has to match only the three couplings
\begin{align}
\label{eq:matching_conditions}
	&\alphatm(\mu_2)=\alphatM^2(\mu_2)\,,&
	&\alpha_\lambda(\mu_2)=\alpha_H(\mu_2)\,,&
	&\alphaV(\mu_2)=\alpha_\nu^2(\mu_2)\,.
\end{align}

Let us now describe how to obtain such target values.
Starting from the scale $M_Z$, we first solve the renormalization group flow of the SM up to the first matching scale $\mu_1=M_\psi=1$ TeV.
In other words we integrate the SM $\beta$-functions given in \Appref{app:smbeta} with boundary conditions for the SM couplings provided by their experimental values, c.f.~\Tabref{tab:IC} or \Citeref{Tanabashi:2018oca}.
At the scale $\mu_1$ the vector-like fermions of the \LVG\ model become dynamical such that,
from this scale on up to $\mu_2=\text{\MGUT}$, we integrate the $\beta$-functions in Eqs.~(\ref{b1}--\ref{bl}), with boundary conditions at $\mu_1$ given by the values obtained from the previous integration.
Clearly, at the scale $\mu_1$, there is one free parameter, namely the value $\alphaV(\mu_1)$ for the beyond-the-SM vector-like Yukawa coupling.
For any values of the latter coupling at $\mu_1$, there will be a set of values for $\alphat$, $\alpha_\lambda$ and $\alphaV$ at $\mu_2$ which defines the IR target for the \LVG\ GUT model.
The value $\alphaV(\mu_1)$ has thus to be fine tuned in order for the matching conditions in \Eqref{eq:matching_conditions} to hold, as we are going to explain in the following.

\subsection{The matching procedure}

Given the IR target $\{\alphatm(\mu_2),\alpha_\lambda(\mu_2),\alphaV(\mu_2)\}$, we have then search for the existence of those trajectories emanating from the fixed points in \Tabref{tab:lvgfps-all} which can be connected to the target itself.
In order to do so, we have integrated the $\beta$-functions for the \LVG\ GUT model given by Eqs.~(\ref{bggut}--\ref{eq:beta-lambda-HSigmap}),
starting from a point infinitesimally close to the selected fixed point and letting the system to flow down to the IR scale $\mu_2=M_\Psi=\text{\MGUT}$.
The initial point of the renormalization group flow is then varied until the trajectory hits, whenever possible, the IR target.

Some comments are in order.
The initial point of the RG flow should belong to the UV critical surface in order to guarantee that the flow towards the UV ends at the considered fixed point, say for example FP$_1$.
This critical surface can be approximated, in the neighborhood of FP$_1$, by its tangent space at FP$_1$ which is defined as the space spanned by the relevant directions at that point.
Of course this approximation is more accurate if the starting point  is closer to  the fixed point; an infinite numerical precision
 would be required in order to lay exactly on the critical surface and a fine tuning problem is always present while flowing towards the UV.
In other words, starting from the IR target there will usually be a positive RG time $t=\log(\mu/\mu_2)>0$ where the numeric integration breaks down entailing the fact that the RG trajectory is repelled away from the critical surface due to nonzero fluctuations along the irrelevant directions.
This is precisely the reason why it is preferable to start the flow in a neighborhood of a fixed point and then flow down to the GUT scale.
This guarantees the fact that the physical trajectories are attracted to the critical surface in the IR.

Let us discuss, as a representative case,  the fixed point FP$_1$ of \Tabref{tab:lvgfps-all} together with the corresponding eigenvalues in \Tabref{tab:lvgfps-all-eigenvalues}.
The tangent space of the critical surface at FP$_1$ is two-dimensional and spanned by the eigenvectors $v_{\rm i}^{(8)}$ and $v_{\rm i}^{(9)}$ associated, respectively, to the negative eigenvalues $\theta_8$ and $\theta_9$ of the stability matrix $M_{\rm ij}$.
Any point on this tangent space can thus be parameterized as
\begin{align}
\label{eq:linear-perturbation-around-alpha*}
	\alpha_{\rm i}&=\alpha_{\rm i}^*+M_{\rm ij}S_{\rm ik}\left(\epsilon^{(8)}z_{\rm k}^{(8)
	}+\epsilon^{(9)}z_{\rm k}^{(9)}\right)
	=\alpha_{\rm i}^*+M_{\rm ij}S_{\rm ik}\left(\epsilon^{(8)}\delta_{\rm k,8}+\epsilon^{(9)}\delta_{\rm k,9}\right)\nn\\
	&=\alpha_{\rm i}^*+\theta_8\epsilon^{(8)}v_{\rm i}^{(8)}+\theta_9\epsilon^{(9)}v_{\rm i}^{(9)}\equiv\alpha_{\rm i}^*+\delta\alpha_{\rm i}\,,
\end{align}
where $\epsilon^{(8)}$ and $\epsilon^{(9)}$ are infinitesimal parameters.
In particular the eigenvector $v_{\rm i}^{(8)}$ is pointing in the $\alphatM$-direction whereas $v_{\rm i}^{(9)}$ induce a displacement along all scalar directions.
The freedom of choosing $\epsilon^{(8,9)}$ allow us to match two couplings with the IR target, namely $\alphatM$ and $\alpha_H$.

Let us consider first the behavior of the top-Yukawa coupling.
Given a positive, however small, displacement $\delta\alphatM(\mu_0)>0$ at a certain initialization scale $\mu_0\gg\mu_2$, the top-Yukawa coupling increases while decreasing the energy scale $\mu$ and, eventually, it crosses its IR target value at a scale $\mu_2'=\mu_0\E^{-t'}$.
On the other hand, given an infinitesimal small, but negative, displacement $\delta\alpha_H(\mu_0)<0$, the quartic Higgs self-interaction coupling decreases while decreasing the scale $\mu$ and, eventually, it crosses its IR target value at a scale $\mu_2''=\mu_0\E^{-t''}$.
Usually the two scales $\mu_2'$ and $\mu_2''$ are different.
Nevertheless, fixing $\epsilon^{(9)}$ while varying $\epsilon^{(8)}$ (or {\it vice versa}), it is possible to fine-tune the initial conditions $\delta\alphat(\mu_0)$ and $\delta\alpha_H(\mu_0)$ such that $\mu_2'=\mu_2''$ to any arbitrarily chosen numerical precision.
Since the initial scale $\mu_0$ is not {\it a priori} fixed, it is legit  to impose $\mu_2'=\mu_2''=\mu_2$ such that,
the fine-tuned initial conditions $\delta\alphatM(\mu_0)$ and $\delta\alpha_H(\mu_0)$ correspond to the correct values at the scale $\mu_0=\mu_2\E^{t'}=\mu_2\E^{t''}$ required in order to match $\alphatM$ and $\alpha_H$ to their IR values in \Eqref{eq:matching_conditions}.

\begin{figure}[t!]
	\begin{center}
	\includegraphics[width=0.85\columnwidth]{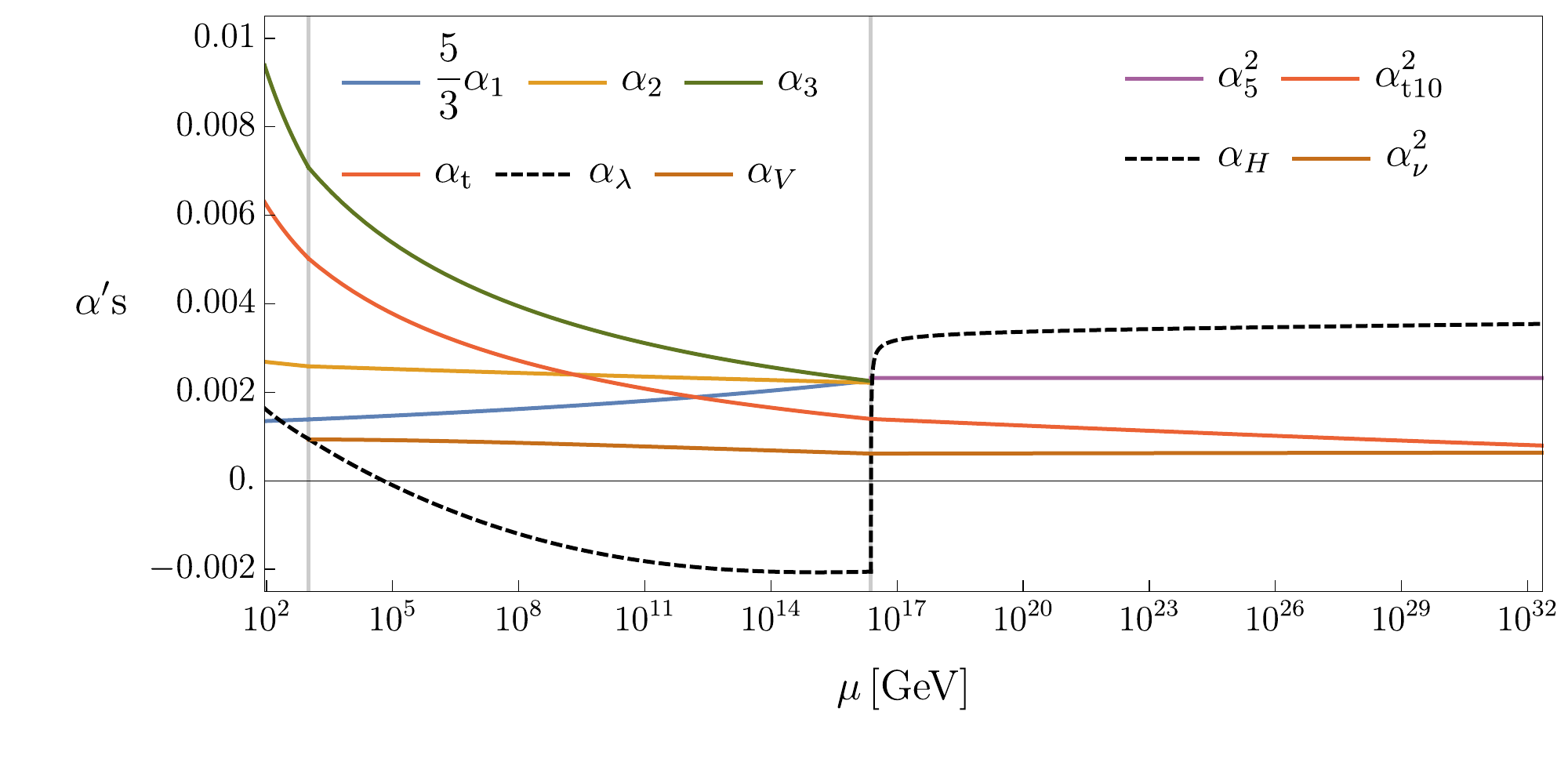}
	%
	\vspace{-10pt}
	\caption{\small
		The renormalization group flow  for the \LVG\ model approaching the fixed point FP$_1$ in the UV limit $\mu\to\infty$.
		Below the GUT scale (at about $10^{16}$ GeV and highlighted by the vertical line in the middle), the gauge couplings $\alpha_i$, the top-Yukawa couplings $\alphatm$ and the Higgs quartic self-interaction are represented with boundary conditions given at the $M_Z$ scale.
		The vector-like fermion fields $\psiL$, $\psiV$ and $\psiG$, with multiplicity $\Nf=1$, enter in the dynamics at the energy scale of 1 TeV (highlighted by the gray vertical line on the left).
		The value for the beyond-the-SM Yukawa coupling $\alphaV$ at 1 TeV is a free parameter to be fine tuned in order to satisfy the matching conditions in \Eqref{eq:matching_conditions}.
		Above the GUT scale, the $\SU(5)$ gauge group is restored and further vector-like fermion fields are included:
		in particular there are 3 flavors of $\Psi_{24}$ and 4 flavors of $\Psi_{5}$.
		The gauge coupling $\alphaM^2$, the top-Yukawa coupling $\alphatM^2$, the Higgs coupling $\alpha_H$ and the vector-like Yukawa coupling $\alpha_\nu^2$ are represented.
			}
	\label{fig:ABGUTrun}
		\end{center}
\end{figure}

There is a technical difficulty regarding the latter fine-tuning procedure.
The RG flow for the quartic Higgs self-interaction $\alpha_H(\mu)$ is such that close to the energy scale $\mu_2''$
it varies very fast due to the presence of an IR singularity below, yet very close to, the scale $\mu_2''$.
At this singular point the quartic Higgs self-interaction diverges to infinitely negative values.
In other words, $\alpha_H(\mu)$ remains very much close to its fixed point value until the energy scale approaches $\mu_2''$ and $\alpha_H(\mu)$ starts to decreases very fast towards the singularity.
It is in this running towards the singularity that the RG trajectory of $\alpha_H(\mu)$ is intercepted and stopped at exactly the scale $\mu_2''$ where $\alpha_H$ equals the (negative) value of the IR target.
Because of the condition $\mu_2''=\mu_2(=\mu_2')$ we can claim that this singularity does not encode a physical inconsistency in our \LVG\  model because it occurs below the GUT scale where, indeed, a different system of differential equations holds.
The only drawback of the presence of this (alleged) singularity below the GUT scale is that, in order to satisfy the equality $\mu_2''=\mu_2=\mu_2'$, an high degree of fine-tuning for the initial conditions $\alphatM(\mu_0)$ and $\alpha_H(\mu_0)$ is required.
In other words, by an appropriate choice for the initial conditions on the (tangent space of the) critical surface at the $\mu_0$ scale, it is possible to move the singularity of $\alpha_H$ below the GUT scale.
This singularity thus becomes physically not worrisome as below the GUT scale another system is considered.

Subsequently, one has to read off the value $\alpha_\nu^2(\mu_2)$, which usually does not coincide with $\alphaV(\mu_2)$.
In order to match this last condition, we have exploited the freedom of choosing different boundary condition for $\alphaV(\mu_1)$.
Varying this latter value one falls into one of these cases $\alpha_\nu^2(\mu_2)\gtrless\alphaV(\mu_2)$ such that a simple bisection algorithm allows to fine-tune the condition $\alpha_\nu^2(\mu_2)=\alphaV(\mu_2)$ to any arbitrary chosen numerical precision.
Notice that for any different value of $\alphaV(\mu_1)$, the above procedure of fine-tuning the couplings $\alphatM$ and $\alpha_H$ has to be repeated, increasing the numerical effort required to satisfy all conditions in \Eqref{eq:matching_conditions}.

Once that all three conditions in \Eqref{eq:matching_conditions} have been satisfied, we can finally plot the full RG flow from the $M_Z$ scale up to the considered fixed point.
As an example, we present how the fixed point FP$_1$ can be connected with the physics at the scale $M_Z$ in \Figref{fig:ABGUTrun} .
At the interface between the two models, that is at the GUT scale, two RG flows described by two sets of first order differential equations have to be matched.
Because of this first order structure, the RG trajectories are only required to be continuous; however, no constraint on the first derivatives, namely the values of the $\beta$-functions at the GUT scale, should be imposed.
This is particular evident in the RG trajectory of the Higgs self-interaction coupling.
The small discontinuity in the gauge couplings is due to the finite correction induced by the one-loop matching condition given by \Eqref{eq:Hall-matching-eq}.

Let us emphasize the fact that the low-energy value for the beyond-the-SM Yukawa coupling $\alphaV$ at the scale $M_\psi=\text{1 TeV}$ represents a physical prediction.
This prediction is the result of the matching conditions at the interface between the two systems below and above the GUT scale.

The above method can be applied to any fixed point in \Tabref{tab:lvgfps-all}.
Quite in general, the latter \Eqref{eq:linear-perturbation-around-alpha*} can be written as
\begin{align}
	\alpha_{\rm i}=\alpha_{\rm i}^*+\sum_{\theta_a<0}\epsilon^{(a)}v_{\rm i}^{(a)}\,,
\end{align}
where the sum is over all the relevant eigen-directions associated to negative eigenvalues and $\epsilon^{(a)}$ are all free parameters.
It seems that the degree of difficulty increases with the number of relevant directions, i.e., the number of free parameters to be fixed.
Nevertheless, we have found that, in order to verify whether a fixed point can be connected with a physical trajectory to the physics at the GUT scale, it is always possible to reduce the number of $\epsilon^{(a)}$ to be fixed to the minimal value of two.
These two parameters are always associated to those directions -- eventually different from $v_{\rm i}^{(8,9)}$ depending on the fixed point---which result in the same RG behavior for the couplings $\alphatM$ and $\alpha_H$ as described above.

In \Tabrefs{tab:lvgfps-all}~and~\ref{tab:lvgfps-all-eigenvalues} we have highlighted in red all possible fixed points which possess, at least, one physical trajectory which hits the IR target $\{\alphat(\mu_2),\alpha_\lambda(\mu_2),\alphaV(\mu_2)\}$ at the GUT scale satisfying the matching conditions in \Eqref{eq:matching_conditions}.
In order to understand why the only highlighted fixed points can be matched with the IR target at the GUT scale, it can be useful to visualize in which directions the linear eigen-perturbations are driving the RG flow.
To this aim, in \Tabref{tab:lvgGUT-linear-perturbations} we have listed with the symbol `` \qedsymbol~'' all those couplings which get perturbed by a nonzero infinitesimal displacement $\epsilon^{(a)}$ along all relevant directions.
It is clear that all fixed points which allow for a match with the IR target at $\mu_2$ share the same feature: both top-Yukawa {\it and} quartic scalar self-interaction couplings have to be perturbed away from their fixed point values.
All other fixed points cannot be matched with the \LVG\ model at the GUT scale because either only the top-Yukawa or the scalar couplings (or neither of them) are perturbed by moving away from the fixed point along the critical surface.

\subsection{The Gau\ss ian fixed point}

We have also investigated the Gau\ss ian fixed point which is present for all values of $N_5$ and  $N_{24}$.
If we further demand the nonabelian gauge coupling $\alphaM$ to be asymptotically free (AF),
then the multiplicities of the vector-like fermions are restricted to be
\begin{align}
\label{eq:multiplicities-for-AF}
	N_{24}=3,N_5\leq 4\,,
	\qquad
	N_{24}=2,N_5\leq 9
	\qquad\text{or}\qquad
	N_{24}=1,N_5\leq 14\,.
\end{align}
Given a model with an AF nonabelian gauge sector, one can try to investigate whether the AF gauge coupling can drive all other couplings towards the Gau\ss ian fixed point.
In order to address this question one can study the {\it quasi-fixed-points}~\Citeref{Gies:2015lia,Gies:2016kkk,Gies:2018vwk,Gies:2019nij} (also called in the literature {\it fixed flows}~\Citeref{Giudice:2014tma} or {\it eigenvalue conditions}~\Citeref{Chang:1978nu,Callaway:1988ya}) for the rescaled couplings
\begin{align}
	\hat{\alpha}_X=\frac{\alpha_X}{\alphaM^2}
	\qquad\text{and}\qquad
	\hat{\alpha}_z=\frac{\alpha_z}{\alphaM}\,,
\end{align}
such that any finite quasi-fixed-point $(\hat{\alpha}_X^*,\hat{\alpha}_z^*)$ represents a specific trajectory along which the UV behavior of the scalar couplings $\alpha_X$ and the Yukawa couplings $\alpha_z$ is locked to follow the AF gauge coupling $\alphaM$.

Among all possible AF scenarios for $\alphaM$ given in \Eqref{eq:multiplicities-for-AF}, we have found the existence of quasi-fixed-points $(\hat{\alpha}_X^*,\hat{\alpha}_z^*)$ only for the combinations
\begin{align}
\label{eq:multiplicities-for-QFP}
	N_{24}=3, N_5=(1,2,4).
\end{align}
Yet, among these possibilities, we have found that none of the corresponding quasi-fixed-points provide viable trajectories which can be matched with the IR target $\{\alphatm(\mu_2),\alpha_\lambda(\mu_2),\alphaV(\mu_2)\}$ at the GUT scale.
In other words, within our \LVG\ GUT model we have found that no total AF trajectories can be found,
thus rendering the interacting fixed points which allow for a matching at the GUT scale even more special.

The vector-like fermion content corresponding to $N_{24}=3,$ $N_5=3$ has been discussed in \Citeref{Giudice:2014tma}, where the authors presented the existence of quasi-fixed-points and argued that realistic total AF GUT models can be constructed.
The reason for this disagreement might  come from the fact that our Yukawa sector is simpler than the one considered in \Citeref{Giudice:2014tma}.
For example, we do not consider the Yukawa interaction term among the vector-like fermion $\Psi_5$ and the SM GUT field $\chi_{\bar 5}$ (exchanging an adjoint scalar $\Sigma$), which is precisely the Yukawa coupling (together with the top-Yukawa) acquiring a nontrivial quasi-fixed-point.
In the light of these observations regarding the Gau\ss ian fixed point, it would be interesting to study how the interacting fixed points presented in \Tabref{tab:lvgfps-all} change after including these terms, in particular those for which a matching with the SM at the GUT scale is possible.

\begin{table}[t!]
	\centering
	\includegraphics[width=0.65\columnwidth]{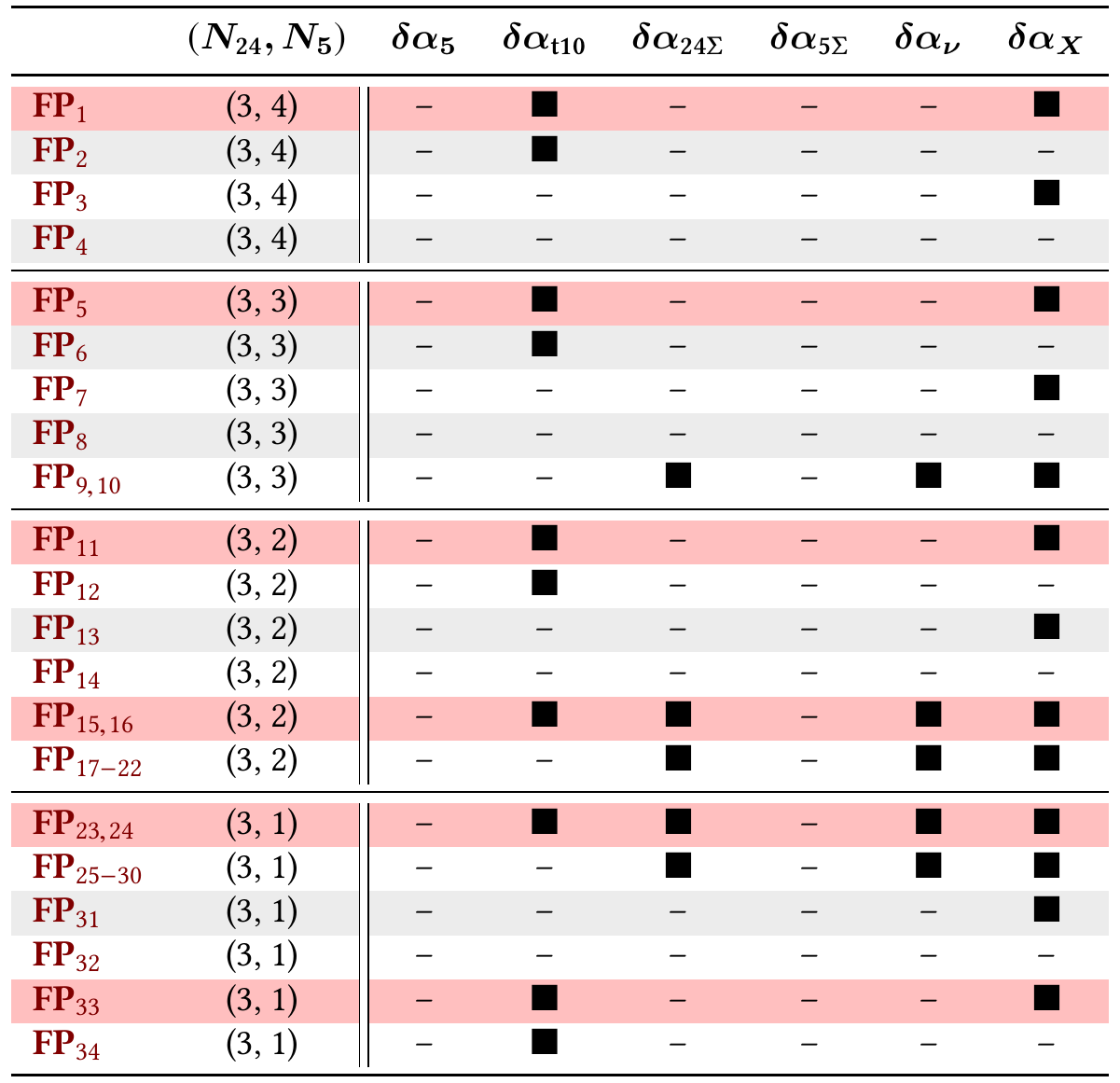}
	\caption{\small
		Schematic representation of those couplings which are perturbed away from the fixed point values by a nonzero infinitesimal displacement $\epsilon^{(a)}$ along all relevant ($\theta_a<0$) directions $v^{(a)}$ spanning the tangent space of the critical surface at the corresponding fixed point.
	}
	\label{tab:lvgGUT-linear-perturbations}
\end{table}

\section{Another minimal extension: the 2U2Q model}
The work of \Citeref{Giudice:2004tc} classifies all minimal  SM extension with vector-like fermions at the TeV scale in which good coupling unification is obtained. Among those, another relevant choice is the \underline{\UUQQ\ model} that represents a ``4th-generation'' scenario. The other extensions seem more far-fetched as they do not lend themselves to a natural physics interpretation.

In the case of the \UUQQ\ model, the ``\Q'' and ``\UU'' labels stand for fields that are vector-like fermion multiplets  transforming, respectively, under the $({\bf 3}, {\bf 2}, 1/6)$ and $({\bf 3}, 1, 2/3)$ representations of the SM, in analogy to the SM quark doublets and up singlets. The \UUQQ\ model corresponds to adding to the SM two $\psiQ$ and two $\psiU$ multiplets at the TeV scale, with quantum numbers shown in \Tabref{tab:2Q2U}.
\begin{table}[t!]
\begin{center} 
\vspace{0.2cm}
\bgroup
\def\arraystretch{1.3}
\begin{tabular}{|c|c|c|c|c|}
\hline
{\bf Fields} & $\SU(3)_\rmc$ & $\SU(2)_\rmL$ & $\U(1)_\rmY$ & $\Nf$ \cr
 \hline 
 $\psiQ$ &  $\mathbf{3}$ & $\mathbf{2}$ & 1/6 & 2\cr
 $\psiU$ &  $\mathbf{3}$ & 1 & 2/3 & 2\cr		
\hline					
\end{tabular}
\egroup
\caption{\small Quantum numbers and multiplicities of the vector-like fermions in the \UUQQ\ model.} 
\label{tab:2Q2U}
\end{center}
\end{table}
The Lagrangian of the \UUQQ\ model then reads
\bea
\label{2Q2Ulag}
{\Lagr}_{\rm 2U2Q}={\Lagr}_{\rm SM}+\overline{\psi}_{Q}^I\I \slashed D \psiQ^I+\overline{\psi}_{U}^J\I \slashed D \psiU^J-\MQ\overline{\psi}_{ Q}^I\psiQ^I
-\MU\overline{\psi}_{ U}^J\psiU^J
-\yQ\sum_{I,J} \overline{\psi}_{ Q}^I \psiU^J H^c + {\rm h.c.}
\eea
where $I,J=(1,2)$ and 
$H^c = \I \sigma_2 H^*$ is the charge conjugated Higgs field.
$\yQ$ is the coupling of the  new Yukawa interaction involving the Higgs and the vector-like fermions.
This term breaks the flavor symmetry to a diagonal subgroup $\SU(2)_Q \times \SU(2)_U \to \SU(2)_D$ , where the ``\Q'' and ``\UU'' vector-like fermions rotate with the same transformation.
At the unification scale, the vector-like fermions of the \UUQQ\ model in \Tabref{tab:2Q2U} are embedded into proper multiples of $\SU(5)$: we assume that the fields $\psiQ$ and $\psiU$ get embedded into vector-like fermions fields $\Psi_{10}$, with multiplicity $N_{10} \geq 2$, transforming under the antisymmetric representation~\footnote{
	The antisymmetric representation of $\SU(5)$ can be decomposed under the SM gauge group as follows~\cite{SLANSKY19811}: ${\bf 10}=({\bf 3},{\bf 2},1/6)\,\oplus\,({\bf \bar 3},1,-2/3)\,\oplus\,(1,1,1)$.}
of $\SU(5)$.
\begin{table}[t!]
\begin{center} 
\label{tab:2Q2UGUT}
\vspace{0.2cm}
\bgroup
\def\arraystretch{1.3}
\begin{tabular}{|c|c|c|}
\hline
{\bf Fields} & $\SU(5)$ & $\Nf$ \cr
 \hline 
 $\Psi_{10}$ &  $\mathbf{10}$ & $N_{10}$\cr
\hline					
\end{tabular}
\egroup
\caption{\small Quantum numbers and multiplicities of the vector-like fermions in the \UUQQ\ GUT model.} 
\end{center}
\end{table}
The Lagrangian of the  \UUQQ\ GUT model is
\begin{align}
	\Lagr_\mathrm{2U2Q}^\mathrm{GUT}=\Lagr_\SM^\mathrm{GUT}
	+\frac{1}{2}\Tr\left(\overline{\Psi}_{10}\I\slashed{D}\Psi_{10}\right)
	-\frac{1}{2}M_{10}\Tr\left(\overline{\Psi}_{10}\Psi_{10}\right)
	+\frac{y_{10}}{8}\sum_{I,J}\varepsilon_{5}\overline{\Psi}_{10}^{cI}\Psi_{10}^J\Phi +\mathrm{h.c.}
	+y_{10}'\sum_{I,J}\Tr\left( \overline{\Psi}_{10}^I\Sigma\Psi_{10}^J\right)\,,
\label{eq:Lagrangian-AGUT-U+Q}
\end{align}
where $I,J=(1,2,\dots,N_{10})$ are the flavor indices for the $\Psi_{10}$ field.
Let us notice that not all possible $\SU(5)$ invariant Yukawa-terms have been considered with the given matter field content.
For the sake of simplicity we have not considered, for example, Yukawa interactions between $\Psi_{10}$ and $\chi_{10}$ or $\chi_{\bar 5}$.
Therefore no Yukawa interactions between the SM fermions and the extra vector-like fermions are retained.

Can this choice of vector-like fermions give rise to an asymptotically safe $\SU(5)$ GUT model? The answer is no.
We computed the $\beta$-functions of this model and studied its fixed points.
We found that, for any value of $N_{10} \geq 2$, the only viable fixed point is the Gau\ss ian one;
yet no good matching with the SM at low energy is possible.
For completeness we give the $\beta$-functions of the \UUQQ\ model in \Appref{app:su52U2Qbetas}.

\section{Outlook}

The goal of having an asymptotically safe extension of the Standard Model is a powerful motivation in searching for physics beyond the SM. It provides a guiding principle that still remains viable  after others, like supersymmetry or compositeness, are waning.

The simplest way to turn the Landau pole of the SM into a fixed point is through a GUT scenario.
The addition of few lepton-like fermions at the 1 TeV scale makes a SU(5) GUT unification of the SM gauge couplings possible and 
consistent with all experimental constraints. We consider what we dubbed the LVG model, which contains the same features of a split SUSY scenario, and   the \UUQQ\ model, which has the features of a 4th generation scenario.

For the \LVG\ model, we  found a GUT embedding that  has a fixed point for the gauge coupling that is interacting---as opposed to the more familiar case of the asymptotically free limit common to all unbroken nonabelian gauge models. We find this an interesting feature. It could perhaps be tested in cosmology, for instance, in physics around and above the GUT phase transition. We also check the \UUQQ\ model but no matching to the SM can be found  in this case.

We are aware that the \LVG\ model as it stands cannot  yet be considered completely satisfactory. 

For one thing, the known problem of the stability of the Higgs potential remains, as it does in the SM. The scale at which the potential crosses to negative values for the coefficient of the quartic term is about  the same  as in the SM, as computed at the one-loop order. It is known~\cite{Isidori:2001bm,EliasMiro:2011aa,Degrassi:2012ry} that this value increases as higher loop orders are computed and included, and  we expect  the same to happen in the case of the LVF model. The model is metastable but  the time scale for its decay is longer than the age of the Universe.

Moreover, fine tuning is required in order to split the masses of scalars and fermions belonging to the same $\SU(5)$ representations as we go to low energy. This is an outstanding problem of all GUT models---indeed, the very  motivation for the original naturalness requirement~\cite{Gildener:1976ai, Weinberg:1978ym,Gildener:1979dd}---to which we have not attempted a solution.

These shortcomings notwithstanding---the asymptotically safe extension of the SM model represented by the \LVG\ model is noteworthy. The theory is UV complete. All the couplings of the model are perturbative and remain so along the entire renormalization group flow up and beyond the Planck scale.  The existence of such a model  is highly non trivial as shown by the lack of fixed points matchable to the SM for models without  unification~\cite{Barducci:2018ysr} or other choices of the vector-like fermion content---to wit, the \UUQQ\ model.

\begin{acknowledgements}
       {\small We thank R.~Percacci and R.~Sondenheimer for valuable discussions. MF is affiliated to the Physics Department of the University of Trieste, the \textit{Scuola Internazionale Superiore di Studi Avanzati} (SISSA) and the Institute for Fundamental Physics of the Universe (IFPU), Trieste, Italy---the support of which is  acknowledged. AT and AU would like to thank ACRI (Associazione di Fondazioni e Casse di Risparmi Spa) and INFN for financial support through the Young Investigator Training Program 2018.
       AU acknowledges support by the DFG under Grants No.  398579334 (Gi328/9-1) and No 396639009 (ZA 958/2-1).
 }
\end{acknowledgements}

\appendix
\section{The SM ${\mathbf{\beta}}$-functions in the \textsc{211-scheme}}
\label{app:smbeta}
In the  \textsc{211-scheme}, the $\beta$-functions of the SM gauge couplings are computed at two-loop, while the $\beta$-functions of the top-Yukawa and Higgs quartic coupling are computed at one-loop.
They are given by (see for example \Citeref{LuoII})
\begin{align}
\beta_1^{\rm SM,NLO}&=\left(\frac{41}{3}+\frac{199}{9}\alpha_{1}+9\alpha_{2}+\frac{88}{3}\alpha_{3}-\frac{17}{3}\alphat \right)\alpha_{1}^{2}\,,\\
\beta_1^{\rm SM,NLO}&=\left(-\frac{19}{3}+3\alpha_{1}+\frac{35}{3}\alpha_{2}+24\, \alpha_{3}-3\,\alphat \right)\alpha_{2}^{2}\,,\\
\beta_1^{\rm SM,NLO}&=\left( -14+\frac{11}{3}\alpha_{1}+9\, \alpha_{2}-52\, \alpha_{3}-4\, \alphat \right)\alpha_{3}^{2}\,,\\
\beta_\rmt^{\rm SM,LO}&=\left( -\frac{17}{6}\alpha_{1}-\frac{9}{2}\alpha_{2}-16\, \alpha_{3}+9\, \alphat \right)\alphat\,,\\
\label{eq:beta-lambda-SM}
\beta_\lambda^{\rm SM,LO}&=12\alpha_{\lambda}^2 	-3\left(\alpha_1+3\alpha_2-4\alpha_\rmt\right)\alpha_\lambda + \frac{3}{4}\left(\alpha_1^2+2\alpha_1\alpha_2+3\alpha_2^2\right)-12\alphat^2\,.
\end{align}
\section{Vector-like fermions contributions to the SM ${\mathbf{\beta}}$-functions}
\label{app:vlfbeta}
Consider Dirac vector-like fermions $\psi$, with multiplicity $\Nf$, that belong to the representation $R_{3}$ of $\SU(3)_\rmc$,  $R_{2}$ of $\SU(2)_\rmL$ and have hypercharge $\rmY$. The one and two-loop contributions to the  running of the SM gauge couplings are given by~\cite{MachacekI,MachacekII,MachacekIII,Luo}:
\begin{align}
	 \beta_1^{\rm NLO}&= \frac{1}{2} \left( B_1 +M_1 \alpha_1 + H_1 \alpha_2  + G_1\alpha_3  \right) \alpha_1 \,,\\
	\beta_2^{\rm NLO}&=\frac{1}{2} \left( B_2 +M_2 \alpha_1 + H_2 \alpha_2  + G_2\alpha_3     \right) \alpha_2 \,,\\
	 \beta_3^{\rm NLO}&=\frac{1}{2} \left( B_3 +M_3 \alpha_1 + H_3 \alpha_2  + G_3\alpha_3  \right) \alpha_3 \,,
	%
	%
	%
	%
	%
\end{align}
where
\begin{align}
B_{1}&=\frac{8}{3}\Nf \rmY^{2}d_{R_{2}}d_{R_{3}}\,, &
B_{2}&=\frac{8}{3}\Nf S_{R_{2}}d_{R_{3}}\,,&
B_{3}&=\frac{8}{3}\Nf S_{R_{3}}d_{R_{2}}\,,\\
M_{1}&=8\rmY^{4}\Nf d_{R_{2}}d_{R_{3}}\,,&
M_{2}&=4\Nf S_{R_{2}}d_{R_{3}}\left(2\, C_{R_{2}}+\frac{20}{3}\right)\,,&
M_{3}&=4\Nf S_{R_{3}}d_{R_{2}}\left(2\,C_{R_{3}}+10\right)\,,\\
H_{1}&=8\rmY^{2}\Nf C_{R_{2}}d_{R_{2}}d_{R_{3}}\,, & 
H_{2}&=8\Nf \rmY^{2}S_{R_{2}}d_{R_{3}}\,, &
H_{3}&=8\Nf\rmY^{2}S_{R_{3}}d_{R_{2}}\,,\\
G_{1}&=8\Nf \rmY^{2}C_{R_{3}}d_{R_{2}}d_{R_{3}}\,,&
G_{2}&=8\Nf S_{R_{2}}C_{R_{3}}d_{R_{3}}\,,&
G_{3}&=8\Nf S_{R_{3}}C_{R_{2}}d_{R_{2}}\,. 
\end{align}
The Casimir invariants $C_{R_i}$ and Dynkin indices $S_{R_i}$ are defined in general as
\begin{align}
d_{R_{2}}&=2\ell+1\,, & d_{R_{3}}&=\frac{1}{2}(p+1)(q+1)(p+q+2)\,, \\
C_{R_{2}}&=\ell(\ell+1)\,, &
C_{R_{3}}&=p+q+\frac{1}{3}(p^{2}+q^{2}+pq)\,, \\
S_{R_{2}}&=\frac{1}{3}d_{R_{2}}C_{R_{2}}\,, & S_{R_{3}}&=\frac{1}{8}d_{R_{3}}C_{R_{3}}\,,   
    \label{Invariants}
\end{align}
where $\ell=0,\frac{1}{2},1,\frac{3}{2},\ldots$ denotes the highest weight of $R_{2}$, and $(p,q)$ (with $p,q=0,1,2\ldots$) the weights of $R_{3}$.
For our specific \LVG\ model we have
\begin{align}
	&\psiL:\, \ell=\frac{1}{2},\,p=q=0\,,&
	&\psiV:\, \ell=1,\,p=q=0\,,&
	&\psiG:\, \ell=0,\,p=q=1\,.
\end{align}

\vspace{\baselineskip}

The contributions to the gauge, top-Yukawa and Higgs quartic couplings coming from the vector-like fermion Yukawa coupling $\alpha_V$,
as well as the contribution to the $\beta$-function $\de_t\alpha_V$ itself, have been computed using the results in \Citeref{XiaoLuo:2003}.
Let us consider two Dirac vector-like fermions $\psi_1$ and $\psi_2$ in two, generically different, representations of the SM gauge group, such that a gauge-invariant Yukawa interaction term can be constructed.
In case $\psi_1$ and $\psi_2$ have multiplicity $N_1$ and $N_2$ respectively, then this interaction term takes the form of
\begin{align}
	- y_{IJ}\, \overline{\psi}_1^I\psi_2^J H + {\rm h.c.}\,,
\end{align}
where $y_{IJ}$ is a complex $N_1\times N_2$ matrix.
For generic representations of $\psi_1$ and $\psi_2$ we were not able to provide general formulas for the contributions of $y_{IJ}$ to the $\beta$-functions for the gauge couplings $\alpha_i$, the top-Yukawa coupling $\alphat$ or for the Higgs quartic self-interaction.
For this reason, an explicit computation of these contributions seems unavoidable.

Let us therefore consider our specif case where $\psi_1=\psiL$ and $\psi_2=\psiV$. We can slightly generalize the vector-like fermion Yukawa interaction in \Eqref{LVGlag} by keeping arbitrary the multiplicities $N_1$ and $N_2$ and assuming that all vector-like fermions interact among each other with the same Yukawa coupling $\yV$.
The contribution to the gauge couplings enters as a two-loop diagram where inside a fermionic loop the Higgs field is exchanged.
This contribution is given by (cf. Eq.~(31) in \Citeref{XiaoLuo:2003})
\begin{align}
\label{eq:VLF-Yukawa-interaction-BGUT}
	\de_t \alpha_i =\dots - \frac{2\alpha_i}{d(\mathcal{G}_i)}\left[C_{R_i}^{(L)}\left(2 d_{R_2}^{(L)}\right)d_{R_3}^{(L)} \mathcal{Y}_{2\rmF}^{(L)}+C_{R_i}^{(V)}d_{R_2}^{(V)}d_{R_3}^{(V)} \mathcal{Y}_{2\rmF}^{(V)}\right] N_1N_2\,,
\end{align}
where $d(\mathcal{G}_i)$ is the dimension of the gauge group $\mathcal{G}_i$ and the superscripts ($L$) and ($V$) refer to the vector-like fermions $\psiL$ and $\psiV$ respectively.
For $i=1$ the Casimir invariants simply reduce to the square of the hypercharges.
The extra factor 2 for the ``\LL'' representation is due to the fact that $\psiL$ is a Dirac-like field while $\psiV$ is a Majorana-like field.
The coefficients $\mathcal{Y}_{2\rmF}$ are the (real) eigenvalues of the matrix product $\mathcal{Y}^a\mathcal{Y}^{a\dagger}$, where $\mathcal{Y}^a$ is the symmetrized matrix of the Yukawa interaction couplings between all Weyl components of the vector-like fermionic fields and the real scalar component $\phi^a$ of $H$
(the construction of the matrices $\calY^a$ will be clarified later in \Appref{app:su5betas}).
For our specific case we have that
\begin{align}
	\mathcal{Y}_{2\rmF}^{(L)}&=3\alphaV^2\,,&
	\mathcal{Y}_{2\rmF}^{(V)}&=4\alphaV^2\,.
\end{align}
Clearly the contribution to the $\beta$-function of the strong gauge coupling is zero since both $\psiL$ and $\psiV$ are singlet under the $\SU(3)_\rmc$ gauge group.

The contribution to the top-Yukawa $\beta$-function comes from the scalar anomalous dimension which has an extra contribution due to the exchange of the vector-like fermions. This extra term reads (cf. Eq.~(33) in \Citeref{XiaoLuo:2003})
\begin{align}
	\de_t\alphat=\beta_\rmt^{\rm SM,LO}  +2\,\alphat\mathcal{Y}_{2\rm S}\,N_1 N_2\,,
\end{align}
where $\mathcal{Y}_{2\rm S}$ is the eigenvalue of the scalar loop matrix $\frac{1}{2}\Tr\left(\mathcal{Y}^{a\dagger}\mathcal{Y}^{b}+\mathcal{Y}^{b\dagger}\mathcal{Y}^{a}\right)=\mathcal{Y}_{2\rm S}\delta^{ab}$.
For our specific case $\mathcal{Y}_{2\rm S}=6\alphaV$.
This justifies also the linear contribution in $\alphaV$ to the $\beta$-function for the Higgs quartic coupling, which is $4\alpha_\lambda\mathcal{Y}_{2\rm S}$.

Similarly, the top-Yukawa contribution to $\de_t\alphaV$ comes from the Higgs anomalous dimension where the top-quark is exchanged. In the latter case, the top contribution to the scalar anomalous dimension is $\mathcal{Y}_{2\rm S}^{\rm top}=3\alphat$, where the factor 3 comes from the color structure.
The gauge contributions to the $\beta$-function for the vector-like Yukawa coupling can be written in terms of the Casimir invariants for the two vector-like representations, such that we can write (cf. Eq.~(33) in \Citeref{XiaoLuo:2003})
\begin{align}
	\de_t\alphaV=15N_1 N_2 \alphaV^2 +2\alphaV \mathcal{Y}_{2\rm S}^{\rm top} - 6\alpha_i\left[C_{R_i}^{(L)}+C_{R_i}^{(V)}\right]\alphaV\,,
\end{align}
where the quadratic contribution in $\alphaV$ depends on the particular form of the matrices $\mathcal{Y}^a$ and is the sum of contributions coming from the fermion anomalous dimension and the renormalization of the operator in \Eqref{eq:VLF-Yukawa-interaction-BGUT}.

Regarding the $\beta$-function for the quartic Higgs coupling, the linear term in $\alphaV$ is due to the scalar anomalous dimension and the quadratic contribution in $\alphaV$ depends again on the particular form of the Yukawa matrices $\mathcal{Y}^a$. We thus obtain (cf. Eq.~(38) in \Citeref{XiaoLuo:2003})
\begin{align}
	\de_t\alpha_\lambda=\beta_\lambda^{\rm SM,LO} 
	+4 N_1 N_2 \mathcal{Y}_{2\rm S}\alpha_\lambda 
	- 48 N_1^2 N_2^2 \alphaV^2\,.
\end{align}

\section{The ${\mathbf\beta}$-functions for the SU(5) LVG model}
\label{app:su5betas}

The gauge and scalar quartic contributions to the $\beta$-functions of the scalar potential in \Eqref{eq:scalar-potential-SU(5)} have been computed at one-loop in \Citeref{Jones2017}.
Given the definitions in \Eqref{eq:def-alphas-AGUT} and defining
$\bar\beta_{X}=\de_t\alpha_{X}$ we have
\begin{align}
\label{eqApp:beta-lambda-Sigma}
\bar\beta_{\Sigma}&=\frac{32}{5}\alpha_{\Sigma}^2
+\alpha_{\Sigma}\left(12\alpha_{\Sigma}'-60 \alphaM^2\right)
+8 \alpha_{H\Sigma}'^{2}+30 \alphaM^4\,,
\\
\label{eqApp:beta-lambda-Sigmap}
\bar\beta_{\Sigma}'&=
32\alpha_{\Sigma}'^2
+\alpha_{\Sigma}'\left(
\frac{94}{5}\alpha_{\Sigma}
- 60 \alphaM^2 \right)
+16\alpha_{H\Sigma}\alpha_{H\Sigma}'
+40 \alpha_{H\Sigma}^2
+\frac{84}{25}\alpha_{\Sigma}^2
+18 \alphaM^4\,,
\\
\label{eqApp:beta-lambda-H}
\bar\beta_{H}&=
18 \alpha_{H}^2-\frac{144}{5}\alpha_{H}\alphaM^2
+96 \alpha_{H\Sigma}^2
+\frac{192}{5}\alpha_{H\Sigma} \alpha_{H\Sigma}'
+\frac{264}{25}\alpha_{H\Sigma}'^2
+\frac{198}{25}\alphaM^4\,,\\
\label{eqApp:beta-lambda-HSigma}
\bar\beta_{H\Sigma}&=
8\alpha_{H\Sigma}^2+
\alpha_{H\Sigma}
\left(12 \alpha_{H}
+\frac{47}{5}\alpha_{\Sigma}
+26 \alpha_{\Sigma}'
-\frac{222}{5}\alphaM^2\right)
+\alpha_{H\Sigma}'\left(
2 \alpha_{H}
+2 \alpha_{H\Sigma}'
+\frac{28}{25}\alpha_{\Sigma}
+\frac{24}{5}\alpha_{\Sigma}'\right)
+\frac{3}{2}\alphaM^4\,,\\
\label{eqApp:beta-lambda-HSigmap}
\bar\beta_{H\Sigma}'&=
\frac{42}{5}\alpha_{H\Sigma}'^2
+ \alpha_{H\Sigma}'\left(
16 \alpha_{H\Sigma}
+\frac{19}{5}\alpha_{\Sigma}
+2 \alpha_{\Sigma}'
+2 \alpha_{H}-\frac{222}{5}\alphaM^2\right)
+\frac{15}{2}\alphaM^4\,.
\end{align}

\subsection{The SU(5) scalar potential}
\label{app:su5betas-scalar-sector}

In order to compute the extra contributions to the latter $\beta$-functions due to the presence of the vector-like fermion Yukawa couplings $\alpha_{5\Sigma}$, $\alpha_{24\Sigma}$ and $\alpha_\nu$, we have made use of the general results of \Citeref{XiaoLuo:2003} (c.f. Eqs.~(38--43) therein).
Following the notation of \Citeref{XiaoLuo:2003}, we have first expanded the scalar fields of our model, namely $\Phi(x)$ and $\Sigma(x)$, in terms of their real scalar components $\phi_a$.
In other words
\begin{align}
	&\Phi(x)=\frac{1}{\sqrt{2}}\begin{pmatrix}
	\phi_1(x)+\I\phi_{6}(x)\,,\,
	\dots\,,\,
	\phi_5(x)+\I\phi_{10}(x)
	\end{pmatrix}^\rmT,
	\label{eq:normalization-Phi}&
	&\Sigma(x)=\sum_{A=1}^{24}\phi_{10+A}(x) T^A\,,
\end{align}
where $T^A$ are the generators of the $\SU(5)$ gauge group, normalized in such a way that $\Tr\left(T^AT^B\right)=\frac{1}{2}\delta^{AB}$.
The quartic scalar potential in \Eqref{eq:scalar-potential-SU(5)} can thus be brought into the following form
\begin{align}
	&V^{\rm quartic}(\Phi,\Sigma)=\frac{1}{4!}\lambda_{abcd}\phi_a\phi_b\phi_c\phi_d\,,&
	&\lambda_{a b c d}=\frac{4!}{\rmP[a,b,c,d]}V\big{|}_{[a,b,c,d]}\,,
\end{align}
where $\lambda_{abcd}$ is a total symmetric rank-4 tensor~\footnote{Generally speaking, for a total symmetric tensor, i.e., symmetric in all its indices, of rank $r$ where all indices can assume $n$ different values, the number of independent components is given exactly by the number of combinations with repetition $C_r^n=(n+r-1)!/(r!(n-1)!)$.
	In our specific case $\lambda_{abcd}$ has $C^{36}_4=66045$ independent components.}
whose entries contain the scalar interactions $\lambda_X$.
$V|_{[a,b,c,d]}$ is the coefficient in front of the quartic operator $\phi_{a}\phi_{b}\phi_{c}\phi_{d}$,
and $\rmP[a,b,c,d]$ is the number of non-equivalent permutations of the set of indices $\{a,b,c,d\}$.
Similarly, all the fermionic fields, namely $\chi_{\bar{5}}$, $\chi_{10}$, $\Psi_5$ and $\Psi_{24}$, have to be expanded in terms of their Weyl left-handed two-components spinors $\psi_i$.
As an explicit example let us consider the vector-like fermion representations
\begin{align}
	\Psi_5&=\begin{pmatrix}
	\xi_1\,,\,\dots\,,\,\xi_5
	\end{pmatrix}^\rmT\,,&
	\xi_i&=\xi_{i,\rmL}+\xi_{i,\rmR}\equiv\psi_{i}-\I\sigma_2\psi_{i+5}^*\,,\\
	\Psi_{24}&=\sum_{A=1}^{24}\xi_{5+A} T^A\,,&
	\xi_{5+A}&\equiv\psi_{10+A}-\I\sigma_2\psi_{10+A}^*\,,
\end{align}
where all the right-handed Weyl components are expressed as the charged conjugation of some Weyl left-handed spinors (notice that the vector-like fermion $\Psi_{24}$ is a Majorana-like fermion).

Given the above decomposition, the Yukawa interaction terms in Eqs.~(\ref{gutlag})~and~(\ref{eq:lagrLVGGUT}) can thus be written as
\begin{align}
	\Lagr^{\rm Yukawa}=-\frac{1}{2}\calY^a_{ij}\psi_i\zeta\psi_j\phi_a+{\rm h.c.}\,,
\end{align}
where $\zeta=\pm\I\sigma_2$ and $\calY^a$ are symmetric Yukawa matrices.

The extra contributions in Eqs.~(\ref{eq:beta-lambda-Sigma}--\ref{eq:beta-lambda-HSigmap}) can be obtained from Eqs.~(40)~and~(41) in \Citeref{XiaoLuo:2003}.
In particular the contributions quadratic in $\alpha_z$ come from the scalars anomalous dimensions and are proportional to the eigenvalues of the scalar loop matrix $\frac{1}{2}\Tr\left(\mathcal{Y}^{a\dagger}\mathcal{Y}^{b}+\mathcal{Y}^{b\dagger}\mathcal{Y}^{a}\right)$.
These eigenvalues are, for the present model,
\begin{align}
	&\calY_{2\rm S}^{H}=	3\alphatM^2+\frac{24}{5}N_5 N_{24}\alpha_\nu^2\,,&
	\calY_{2\rm S}^{\Sigma}=\frac{21}{20}N_{24}\alpha_{24\Sigma}^2+N_5\alpha_{5\Sigma}^2\,.
\label{eq:scalar-eigenvalues}
\end{align}
The quartic contributions in $\alpha_z$ are instead due to a fermionic loop where four fermions are exchanged among the four scalar fields.
This contribution is obtained from
\begin{align}
	\frac{\de_t\lambda_{abcd}}{(4\pi)^2}=\dots -\sum \Tr\left[\calY^a\calY^{b\dagger}\calY^c\calY^{d\dagger}\right]\,,
\end{align}
where the sum is over all 4! permutations of the indices $\{a,b,c,d\}$.

\subsection{The gauge-Yukawa subsystem of the SU(5) LVG model}
\label{app:su5betas-gaugeYukawa-sector}

The $\beta$-functions for the $\SU(5)$ gauge coupling as well as for the Yukawa couplings have also been computed by mean of the general formulas for a generic gauged Quantum Field Theory given in \Citeref{XiaoLuo:2003}.

In particular the $\beta$-functions for the Yukawa couplings have been computed from Eq.~(33) of \Citeref{XiaoLuo:2003}.
The extra contributions to $\de_t\alphatM$ due to the vector-like Yukawa interaction $\alpha_\nu$, as well as the top-contribution to $\de_t\alpha_\nu$, come from the Higgs anomalous dimension, c.f. \Eqref{eq:scalar-eigenvalues}.
The terms proportional to $N_5$ and $N_{24}$ in Eqs.~\eqref{b5gut}~and~\eqref{b24gut} come, instead, from the anomalous dimension of the $\Sigma$ scalar field, c.f. again \Eqref{eq:scalar-eigenvalues}.
The other Yukawa contributions are a nontrivial sum of the fermionic anomalous dimensions and the renormalization of the operator $\calY^a_{ij}\psi_i\zeta\psi_j$.
Therefore an explicit computation of the first two terms in Eq.~(33) of \Citeref{XiaoLuo:2003} is required.

The gauge contributions to the $\beta$-functions of the Yukawa couplings are obtained by computing the Casimir invariants for the different fermionic representations.
For a generic $\SU(N)$ gauge group, the Casimir for the fundamental, antisymmetric and adjoint representations are, respectively,
\begin{align}
	&C_{N}=\frac{N^2-1}{2N}\,,&
	&C_{\rm antisymm}=\frac{(N+1)(N-2)}{N}\,,&
	&C_{\rm adj}=N\,,
\end{align}
such that the gauge contributions to the Yukawa $\beta$-functions for our \LVG\ $\SU(5)$ GUT model are
\begin{align}
\label{eq:beta-alpha_tb}
	\de_t\alphatM&=\dots - 3 \alphaM^2 \left(C_{\bf 10}+C_{\bf 10}\right) \alphatM\,,&
	\de_t\alpha_{5\Sigma}&=\dots - 3\alphaM^2\left(C_{\bf 5}+C_{\bf 5}\right)\alpha_{5\Sigma}\,,\\
	\de_t\alpha_{24\Sigma}&=\dots - 3\alphaM^2\left(C_{\bf 24}+C_{\bf 24}\right)\alpha_{24\Sigma}\,,&
	\de_t\alpha_{\nu}&=\dots - 3\alphaM^2\left(C_{\bf 5}+C_{\bf 24}\right)\alpha_{\nu}\,,
\end{align}
where the sum in parenthesis refers to the sum over the fermions which are exchanged in the one-loop diagrams where a gauge boson is exchanged.

The two-loop $\beta$-function for the $\SU(5)$ gauge coupling has been computed from Eq.~(30) of \Citeref{XiaoLuo:2003}, where the terms proportional to the square of the Yukawa couplings can be written as
\begin{align}
	\de_t \alphaM =\dots - \frac{\alphaM^3}{d(\mathcal{G})}\left[
C_{\bf 10} d_{\bf 10}\calY_{2\rmF}(\chi_{10})+
C_{\bf 5} d_{\bf 5}\calY_{2\rmF}(\Psi_5) (2 N_{5})+
C_{\bf 24} d_{\bf 24}\calY_{2\rmF}(\Psi_{24}) N_{24}
\right]\,,
\end{align}
where $d({\cal G})=24$ is the dimension of the $\SU(5)$ gauge group and $d_{\rmF_i}$, with $F_i=(\bf{5,10,24}$), is the dimension of the different fermionic representations.
Let us notice the presence of an extra factor of 2 in the multiplicity of $\Psi_5$ due to the fact that it is a Dirac-like fermion. 
The eigenvalues of the matrix product $\calY^a\calY^{a\dagger}$ are
\begin{align}
	\calY_{2\rmF}(\chi_{10})&=3\alphatM^2\,,&
	\calY_{2\rmF}(\Psi_{5})&=\frac{12}{5}\alpha_{5\Sigma}^2+\frac{12}{5}N_{24}\alpha_\nu^2\,,&
	\calY_{2\rmF}(\Psi_{24})&=\frac{21}{20}\alpha_{24\Sigma}^2+N_5\alpha_\nu^2\,.
\end{align}

The one-loop contributions to the RG flow of $\alphaM$ is obtained by first computing the Dynkin indices for the different (fermionic and scalar) representations. Generally speaking, given a representation $R$ of a gauge group ${\cal G}$, we have
\begin{align}
	S_{R}=\frac{d_R C_R}{d({\cal G})}\,,
\end{align}
such that for $\SU(N)$ we obtain
\begin{align}
	S_{N}&=\frac{1}{2}\,,&
	S_{\rm antisymm}&=\frac{N-2}{2}\,,&
	S_{\rm adj}&=N\,.
\label{eq:Dynkin-indices}
\end{align}
For our specific $\SU(5)$ GUT model we thus have
\begin{align}
	\de_t\alphaM=-\alphaM^3\left[\frac{11}{3}C({\cal G})
	-\sum_i\left(
	\frac{2}{3}S_{\rmF_i}
	+\frac{1}{6}S_{{\rm S}_i}\right)\right]+\beta_{\alphaM}^{\rm NLO}
\end{align}
where the Casimir for the $\SU(5)$ gauge group is $C({\cal G})=5$ and the sum is over all the fermionic and scalar representations.
Each generation of the SM fermionic sector can be fitted in the representations ${\bf \bar 5}$ and ${\bf 10}$.
On the other hand, the scalar sector of the SM GUT theory is composed of a complex fundamental representation and a real adjoint representation.
Taking into account also the vector-like fermion representations we finally have
\begin{align}
	\de_t\alphaM=-\alphaM^3\left\{\frac{11}{3}C({\cal G})
	-\frac{2}{3}\Bigl[S_{\rmF}(\chi_{\bar{5}})+S_{\rmF}(\chi_{10})\Bigr]N_{\rm g}
	-\frac{2}{3}\Bigl[2N_5S_{\rmF}(\Psi_{5})+N_{24}S_{\rmF}(\Psi_{24})\Bigr]
	-\frac{1}{6}\Bigl[2\,S_{\rm S}(\Phi)+S_{\rm S}(\Sigma)\Bigr]\right\}+\beta_{\alphaM}^{\rm NLO}\,,
\end{align}
where $N_{\rm g}=3$ is the generation number and the extra factor of 2 for the contributions of $\Phi$ and $\Psi_5$ come from the fact that $\Psi_5$ is a Dirac-like fermion and $\Phi$ is composed of two real 5-plets, respectively the real and imaginary parts.
Substituting the values in \Eqref{eq:Dynkin-indices} into the latter equation, we obtain the one-loop contribution of \Eqref{bggut}.
Given the Casimir and Dynkin indices for the different representations, it is straightforward to obtain the two-loop contribution in \Eqref{bggut} from Eq.~(30) of \Citeref{XiaoLuo:2003}.

\section{The ${\mathbf\beta}$-functions for the 2U2Q model}
\label{app:su52U2Qbetas}

Using the definitions in \Eqref{alphasbgut} and the following rescaling for the coupling $\alphaQ$
\begin{align}
	\alphaQ=\frac{\yQ^2}{(4\pi)^2}\,,
\end{align}
the $\beta$-functions of the \UUQQ\ model read
\begin{align}
\de_t \alpha_1&= \beta_1^{\rm SM,NLO}+\left(8+\frac{86}{9}\alpha_1+2\alpha_2+32\alpha_3-\frac{136}{3}\alphaQ\right) \alpha_1^2\,,\\
\de_t \alpha_2&= \beta_2^{\rm SM,NLO}+\left(8+\frac{2}{3}\alpha_1+98\alpha_2+32\alpha_3-24\alphaQ\right) \alpha_2^2\,,\\
\de_t \alpha_3&= \beta_3^{\rm SM,NLO}+\left(8+4\alpha_1+12\alpha_2+152\alpha_3-32\alphaQ\right)\alpha_3^2\,,\\
\de_t \alphat&= \beta_\rmt^{\rm SM,LO}+48\alpha_Q \alphat\,,\\
\de_t \alphaQ&= \left(- \frac{17}{6} \alpha_1 - \frac{9}{2}\alpha_2 - 16 \alpha_3 + 60 \alphaQ + 6 \alphat\right)\alphaQ\,,\\
\de_t \alpha_\lambda &= \beta_\lambda^{\rm SM,LO} + 96\alphaQ \alpha_\lambda - 384\alphaQ^2\,,
\end{align}
where $\beta_i^{\rm SM,NLO}$, $\beta_\rmt^{\rm SM,LO}$ and $\beta_\lambda^{\rm SM,LO}$  are the SM $\beta$-functions previously given in \Appref{app:smbeta}.
The new terms arising from the presence of the extra vector-like fermions $\psiU$ and $\psiQ$ are explicitly shown.
Their contributions to the running of the gauge couplings have been computed using the formulas in \Appref{app:vlfbeta}.
The extra contributions due to the vector-like Yukawa coupling $\alphaQ$ as well as the $\beta$-function of $\alphaQ$ itself have been computed using the results of \Citeref{XiaoLuo:2003} (c.f.~Eqs.~(30,~33)~and~(38) therein).

\vspace{\baselineskip}

In the following we present the $\beta$-functions of the \UUQQ\ GUT model where the vector-like fermions $\psiU$ and $\psiQ$ become embedded into the antisymmetric representation ${\bf 10}$ of $\SU(5)$.
Given $y_{10}$ and $y_{10}'$ the two vector-like Yukawa couplings defining the interactions between $\Psi_{10}$ and the scalar fields $\Phi$ and $\Sigma$ respectively, we define
\begin{align}
	&\alpha_{10} =\frac{y_{10}}{4\pi}\,,&
	&\alpha_{10}' =\frac{y_{10}'}{4\pi}\,,
\end{align}
together with the definitions given in \Eqref{eq:def-alphas-AGUT} for the gauge, top-Yukawa and scalar couplings.
For the sake of simplicity, we assume that the latter Yukawa interactions are diagonal in the flavor indices $I,J=(1,2,\dots,N_{10})$,
such that different flavors of the matter field $\Psi_{10}$ do not interact among each others.
The sum in \Eqref{eq:Lagrangian-AGUT-U+Q} reduces to $\sum_{I}(\dots).$
The RG flow equations for the gauge-Yukawa subsystem are
\begin{align}
\de_t\alphaM&= -\frac{40-6N_{10}}{3}\alphaM^3
-\frac{1184-1074 N_{10}}{15} \,\alphaM^5
-\left(   \frac{9}{2}\alphatM^2+9N_{10}\alpha_{10}^2
+\frac{54}{5}N_{10}\alpha_{10}'^2
\right)\alphaM^3\,,\\
\de_t\alphatM&=\left(
6\alphatM^2 	
+6N_{10}\alpha_{10}^2- \frac{108}{5}\alphaM^2\right)\alphatM\,,
\label{eq:beta-top-U+Q}\\
\de_t\alpha_{10}&=\left[
3\left(1+2N_{10}\right)\alpha_{10}^2+3\alphatM^2
-\frac{6}{5}\alpha_{10}'^2
- \frac{108}{5}\alphaM^2 \right] \alpha_{10}\,,
\label{eq:beta-y_10}\\
\de_t\alpha_{10}'&=\left[\frac{29+15N_{10}}{5}\alpha_{10}'^2 -\alpha_{10}^2
- \frac{108}{5}\alphaM^2\right]
\alpha_{10}'\,.
\label{eq:beta-y_10p}
\end{align}
We conclude the set of $\beta$-functions by listing the running of the scalar couplings.
\begin{align}
\de_t\alpha_{\Sigma}&=\bar \beta_\Sigma+12 N_{10}\left(\alpha_{\Sigma}+\alpha_{10}'^2\right)\alpha_{10}'^2\,,\\
\de_t\alpha_{\Sigma}'&=\bar \beta _{\Sigma}' +12N_{10}\left(\alpha_{\Sigma}'-\alpha_{10}'^2\right)\alpha_{10}'^2\,,\\
\de_t\alpha_{H}&=\bar \beta_H+\left(12\alphatM^2+12N_{10}\alpha_{10}^2\right)\alpha_{H} -12\alphatm^4-24N_{10}\alpha_{10}^4\,,\\
\de_t\alpha_{H\Sigma}&=\bar \beta_{H\Sigma}
+\left(6\alphatM^2+12N_{10}\alpha_{10}^2+6N_{10}\alpha_{10}'^2 \right)\alpha_{H\Sigma} - 4N_{10}\alpha_{10}^2\alpha_{10}'^2\,,\\
\de_t\alpha_{H\Sigma}'&=\bar\beta_{H\Sigma}'
+\left(6\alphatM^2+12N_{10}\alpha_{10}^2+6N_{10}\alpha_{10}'^2 \right)\alpha_{H\Sigma}'-4N_{10}\alpha_{10}^2\alpha_{10}'^2\,,
\end{align}
where the pure gauge e scalar contributions $\bar \beta_\Sigma$, $\bar \beta_\Sigma'$, $\bar \beta_H$, $\bar \beta_{H\Sigma}$, $\bar \beta_{H\Sigma}'$ are given in \Appref{app:su5betas}.

\cleardoublepage
\bibliographystyle{utphys.bst}
\bibliography{safeGUT.bib}

\end{document}